\documentclass[useAMS,usenatbib]{mnras}
\usepackage{graphicx}

\title[On the Astrid asteroid family]
{On the Astrid asteroid family}
\author[V. Carruba]{V. Carruba$^{1}$\thanks{E-mail: vcarruba@feg.unesp.br}\\
$^{1}$UNESP, Univ. Estadual Paulista, Grupo de din\^{a}mica Orbital e
  Planetologia, Guaratinguet\'{a}, SP, 12516-410, Brazil. \\
}

\begin{document}

\date{Accepted ... .  Received 2016 ...; in original form 2016 May 02}

\pagerange{\pageref{firstpage}--\pageref{lastpage}} \pubyear{2016}

\maketitle

\label{firstpage}

\begin{abstract}
Among asteroid families, the Astrid family is peculiar because of its unusual 
inclination distribution.  Objects at $a\simeq$~2.764 au are quite dispersed 
in this orbital element, giving the family a ``crab-like'' appearance.   
Recent works showed that this feature is caused by the interaction of the 
family with the $s-s_C$ nodal secular resonance with Ceres, that spreads the 
inclination of asteroids near its separatrix.  As a consequence, the currently 
observed distribution of the $v_W$ component of terminal ejection velocities 
obtained from inverting Gauss equation is quite leptokurtic, since this 
parameter mostly depends on the asteroids inclination.  The peculiar orbital 
configuration of the Astrid family can be used to set constraints on key 
parameters describing the strength of the Yarkovsky force, such as the 
bulk and surface density and the thermal conductivity of surface material.  
By simulating various fictitious families with different values of these 
parameters, and by demanding that the current value of the kurtosis of the 
distribution in $v_W$ be reached over the estimated lifetime of the family, 
we obtained that the thermal conductivity of Astrid family members should be 
$\simeq$ 0.001 W/m/K, and that the surface and bulk density should be higher 
than 1000 kg/m$^{3}$.  Monte Carlo methods simulating Yarkovsky and stochastic 
YORP evolution of the Astrid family show its age to be $T$ = 140$\pm$30 Myr 
old, in good agreement with estimates from other groups.   Its terminal 
ejection velocity parameter is in the range $V_{EJ}= 5^{+17}_{-5}$~m/s.  Values 
of $V_{EJ}$ larger than 25 m/s are excluded from constraints from the current 
inclination distribution.

\end{abstract}

\begin{keywords}
Minor planets, asteroids: general -- minor planets, asteroids: individual: 
Astrid-- celestial mechanics.  
\end{keywords}
%

\section{Introduction}
\label{sec: intro}

The Astrid asteroid family is characterized by an unusual distribution
in the $(a,\sin{(i)})$ plane, with a dispersion in inclination
of its members at $a \simeq $~2.764~au much larger than that of members
at other semi-major axis.  \citet{Novakovic_2016} recently
showed that this feature of the Astrid family is caused by its interaction
with the $s-s_C$ nodal secular resonance with Ceres.  Asteroid crossing
this resonance are significantly dispersed in inclination, causing
the crab-like appearance of the family.  The unusual distribution 
in inclination of the Astrid family also produces other consequences.
\citet{Carruba_2016} observed that the current distribution of the
$v_W$ component of terminal ejection velocities field computed from
inverting Gauss equation for this family 
is characterized by a leptokurtic distribution,
i.e., a distribution with larger tails and more peaked than a Gaussian.
If we define as kurtosis the ratio of the fourth momenta of a 
distribution with respect to the fourth power of its standard deviation,
that for a distribution of n random variable $x_i$ is given by:

\begin{equation}
k=\frac{\frac{1}{n}\sum_{i=1}^{n}{(x_i-<x>)^4}}{(\frac{1}{n}\sum_{i=1}^{n}{(x_i-<x>)^2)^2}},
\label{eq: kurtosis}
\end{equation}

\noindent  where $<x>=\frac{1}{n}\sum_{i=1}^{n}{x_i}$ is the mean value of the
distribution, then Pearson ${\gamma}_2$ kurtosis is equal to k-3.  Gaussian
distributions are characterized by values of ${\gamma}_2$ equal to 0.
The value of the Pearson ${\gamma}_2$ parameter for the whole Astrid family
is quite high, but is closer to mesokurtic values if asteroids in the
resonant region are excluded.

In this work we investigate what information on key parameters 
describing the Yarkovsky effect, such as the thermal conductivity of material 
on the surface and the mass density, can be obtained by studying the 
orbital diffusion of fictitious members of several simulated Astrid
families.  By checking on what time-scales the current value of
${\gamma}_2(v_W)$ can be reached, and for what values of the parameters
describing the Yarkovsky force, constraints on the allowed range of 
values of these parameters can be, in principle, obtained.
The independent constraints provided by secular dynamics (and from
the current inclination distribution of Astrid members) could then 
be used to estimate the age of the Astrid family with a higher precision
than that available for other families.

\section{Family identification and local dynamics}
\label{sec: fam_ide}

As a first step in our analysis we selected the Astrid family, as identified
in \citet{Nesvorny_2015} using the Hierarchical Clustering Method
(HCM, \citep{Bendjoya_2002}) and a cutoff of 60~m/s.  489 members of the
Astrid dynamical group were identified in that work.  We also selected
asteroids in the background of the family, defined as a box in the
$(a,e,\sin{(i)})$ domain.  We selected asteroids to within the minimum
and maximum values of Astrid proper elements, plus or minus 0.02 au, 0.02,
and 0.02 in proper $a$,$e$, and $\sin{(i)}$, respectively,
with the exception of the maximum values in $a$ that was given by 
the semi-major axis of the center of the 5J:-2A mean-motion resonance.
588 asteroids, 99 of which not members of the Astrid group, were identified
in the background of the family so defined.

\begin{figure*}
\centering
\centering \includegraphics [width=0.85\textwidth]{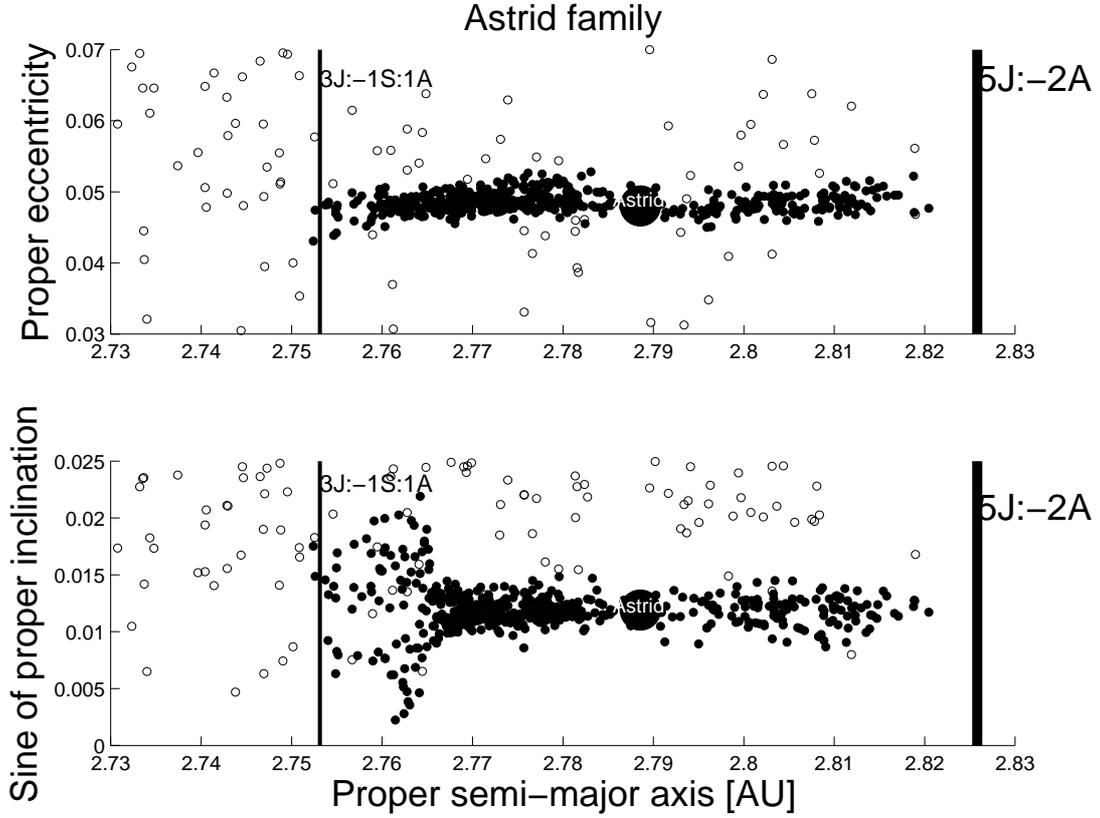}

\caption{A $(a,e)$ (top panel) and $(a,\sin{(i)}$ (bottom panel) projection
of members of the HCM Astrid cluster (489 members, black full dots), 
and of the local background (588 members, black open dots).  Vertical 
lines display the location of the local mean-motion resonances.  The 
orbital location of 1128 Astrid is identified by a large black circle 
and it is labeled.}
\label{fig: Astrid_aei}
\end{figure*}

\begin{figure*}
  \centering
  \begin{minipage}[c]{0.50\textwidth}
    \centering \includegraphics[width=3.4in]{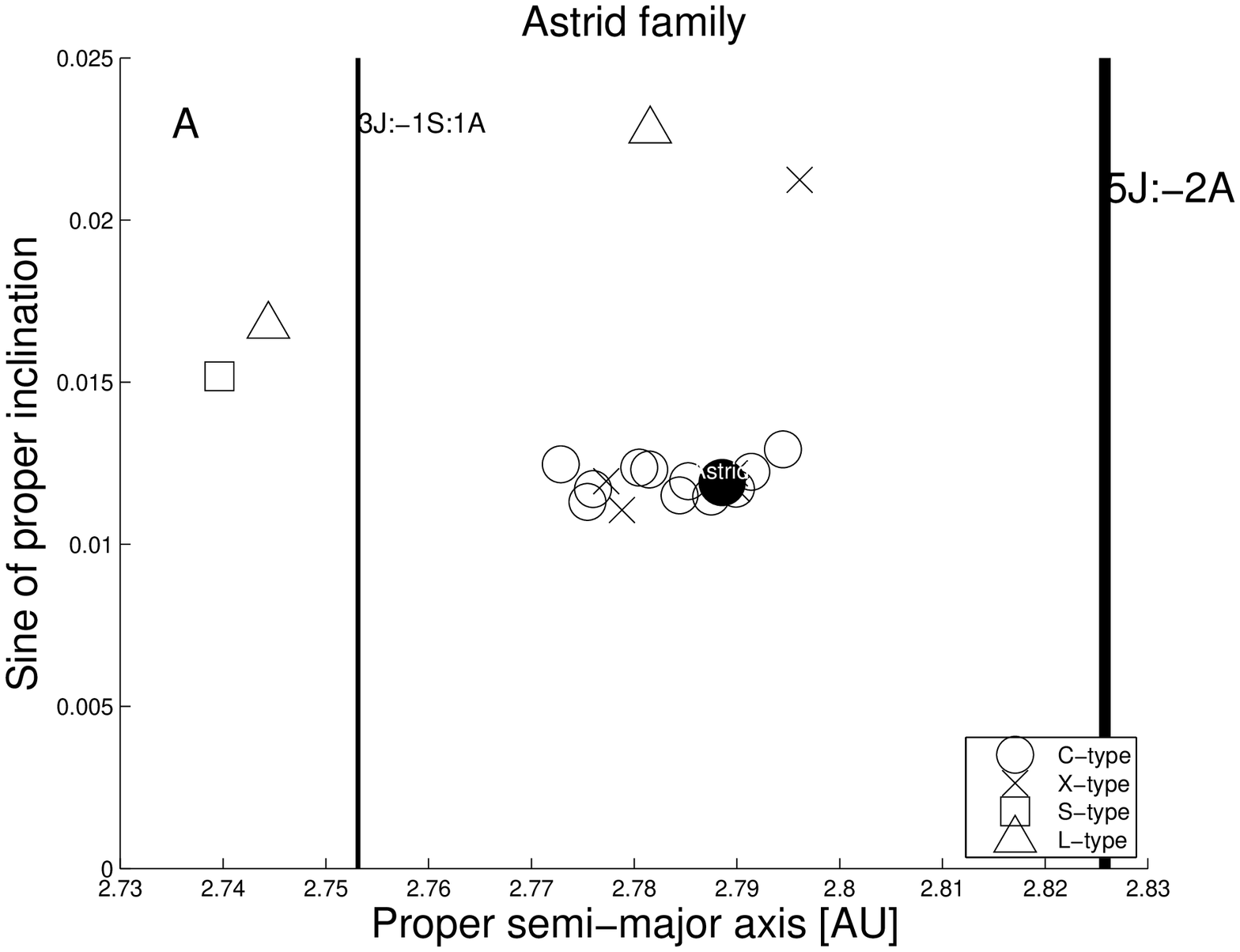}
  \end{minipage}%
  \begin{minipage}[c]{0.50\textwidth}
    \centering \includegraphics[width=3.4in]{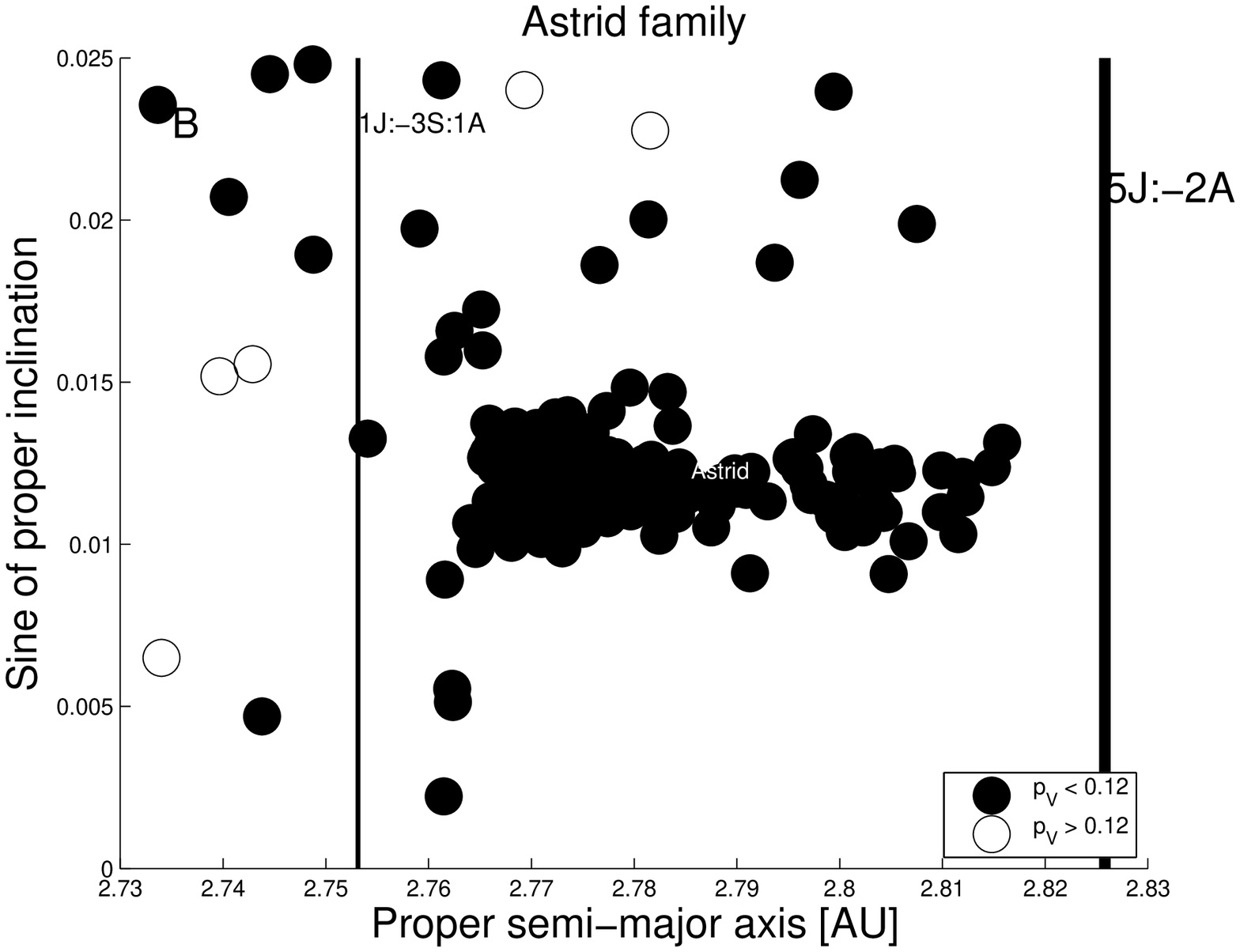}
  \end{minipage}

\caption{An $(a,\sin{(i)}$ projection of the 20 asteroids with taxonomic 
information (panel A) and of the 207 bodies with WISE albedo (panel B).
See figures legends for the meaning of the used symbols.} 
\label{fig: Astrid_sdss_pv}
\end{figure*}

Fig.~\ref{fig: Astrid_aei} displays the orbital location of family members
(black full dots) and local background asteroids (black open dots) in the 
$(a,e)$ (top panel) and $(a,\sin{(i)})$ (bottom panel) domains.   The Astrid
family numerically dominates the population in the local background: 83.1\%
of the asteroids in the region are members of the HCM family.  One can also
notice the spread in $\sin{(i)}$ of Astrid members at $a \simeq 2.765$~au,
caused by the nodal linear secular resonance with Ceres $s-s_C$, as
shown in \citet{Novakovic_2016}. 

We then turned our attention to the physical properties of objects in the
Astrid region.  We checked which asteroids have information in the three 
major photometric/spectroscopic surveys (ECAS (Eight-Color Asteroid 
Analysis, \citet{Tholen_1989}), SMASS (Small Main Belt Spectroscopic Survey, 
\citet{Bus_2002a,Bus_2002b}), and S3OS2 (Small Solar System Objects 
Spectroscopic Survey, \citet{Lazzaro_2004}), in the Sloan Digital Sky 
Survey-Moving Object Catalog data, fourth release (SDSS-MOC4 hereafter, 
\citet{Ivezic_2001}), and in the WISE survey \citep{Masiero_2012}.
Taxonomic information was deduced for the SDSS-MOC4 objects using the 
method of \citet{DeMeo_2013}.  We obtained taxonomic information
for 20 asteroids, while 207 bodies had values of geometric albedo
in the WISE data-set.     Fig.~\ref{fig: Astrid_sdss_pv} displays our
results for these objects.   The Astrid family is a C-complex family,
and C-complex objects dominate the local background: out of 
207 bodies with information on geometric albedo, only 5 (2.4\% of the total)
have $p_V > 0.12$, and are possibly associated with a S-complex composition.
No taxonomic or albedo interlopers were identified in the Astrid HCM
group.

How much the local dynamics is responsible for the current shape of the Astrid
family?  To answer this question, we obtained dynamical maps in the 
domain of proper $(a,\sin{(i)})$ with the method described in 
\citet{Carruba_2010}, based on the theory developed by
\citet{Knezevic_2000}.  We integrated 1550 particles over 20 Myr under 
the gravitation influence of i) all planets and ii) all planets plus Ceres as a
massive body\footnote{The mass of Ceres was assumed to be equal
to $9.39 \cdot 10^{20}$~kg, as determined by the Dawn spacecraft 
\citep{Russell_2015}.} with $SWIFT\_MVSF$, the symplectic integrator
based on $SWIFT\_MVS$ from the {\em Swift} package of \citet{Levison_1994},
and modified by \citet{Broz_1999} to include on line filtering of osculating
elements.  The initial osculating elements of the particles went
from 2.730 to 2.828~au in $a$ and from $1.00^{\circ}$ to $2.45^{\circ}$ in 
$i$.  We used 50 intervals in $a$ and 31 in $i$.  The other 
orbital elements of the test particles were set equal to 
those of Ceres at the modified Julian date of 57200.  

\begin{figure*}
  \centering
  \begin{minipage}[c]{0.50\textwidth}
    \centering \includegraphics[width=3.5in]{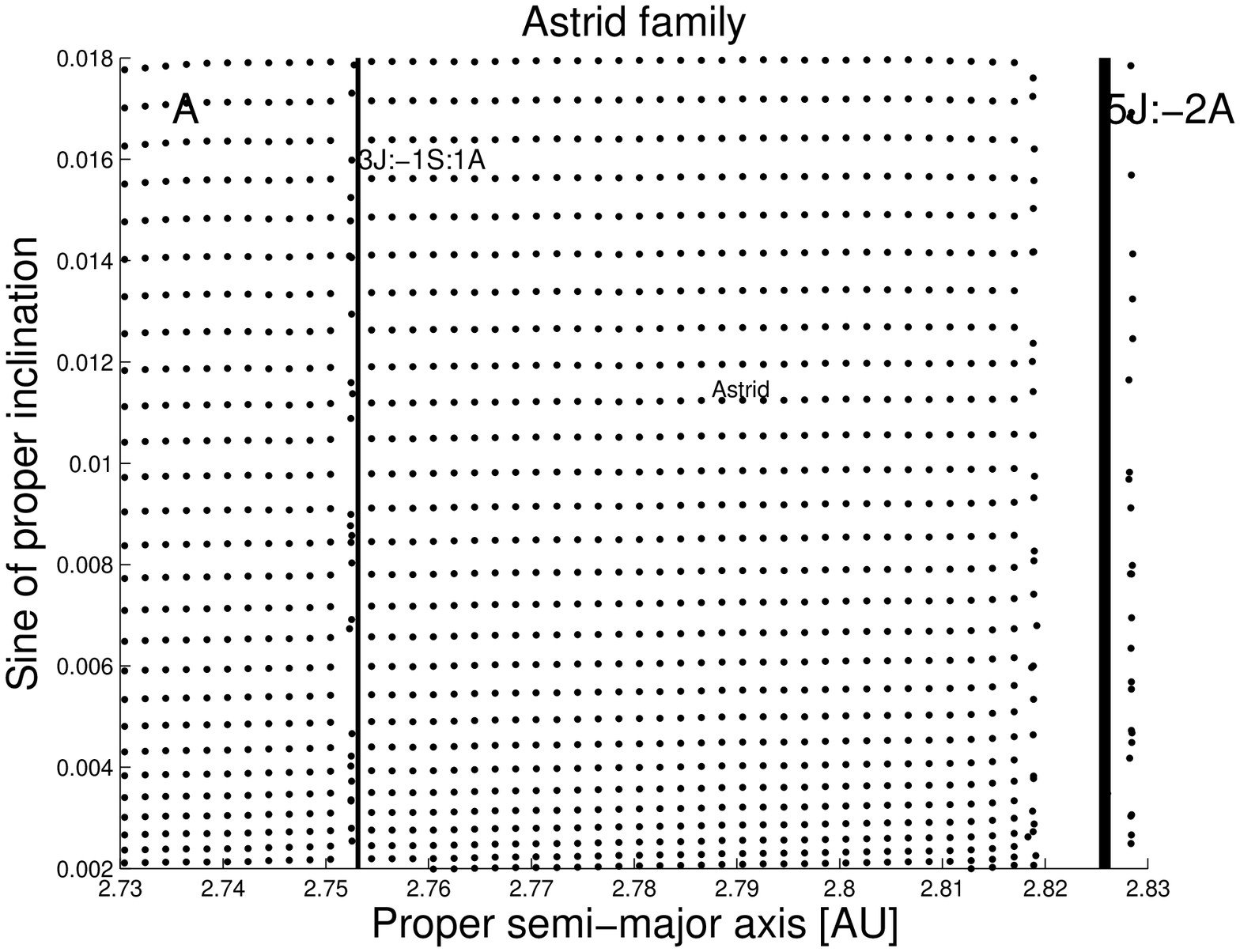}
  \end{minipage}%
  \begin{minipage}[c]{0.50\textwidth}
    \centering \includegraphics[width=3.5in]{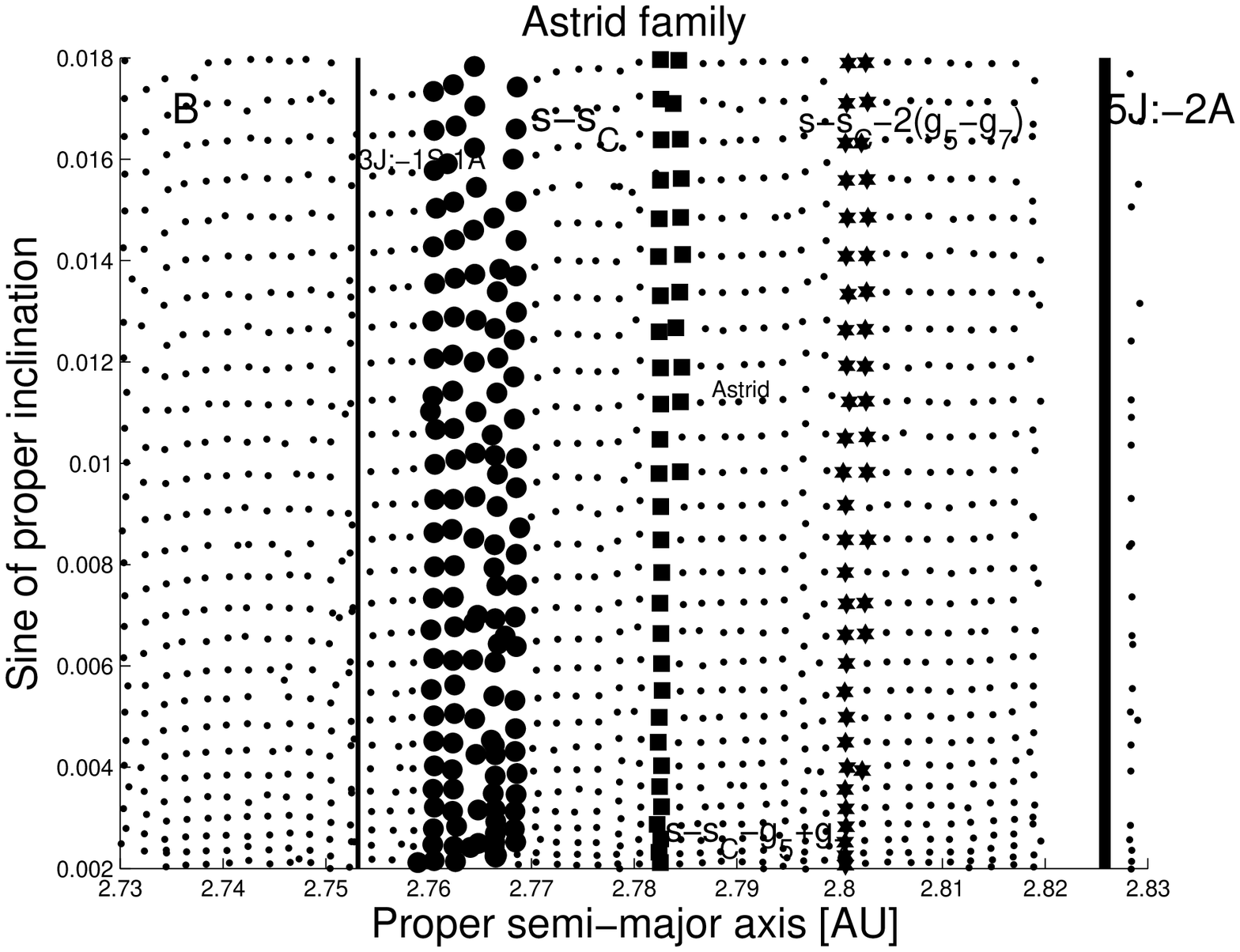}
  \end{minipage}

\caption{Dynamical maps for the orbital region of Astrid obtained by
integrating test particles under the influence of all planets (panel A), 
and all planets and Ceres as a massive body (panel B). Unstable
regions associated with mean-motion resonances appear as vertical  
strips.  Secular resonance appear as inclined bands of 
aligned dots.  Dynamically stable regions are shown as uniformly 
covered by black dots.  Vertical lines display the location of the main 
mean-motion resonances in the area.  Black filled dots in panel B show 
the locations of ``likely resonators'' in the $s-s_C$ secular 
resonance.  Likely resonators in the  $s-s_C-g_5+g_7$ and $s-s_C-2(g_5+g_7)$
resonances are shown as full squares and full hexagons, 
respectively.} 
\label{fig: Maps_ai}
\end{figure*}

Fig.~\ref{fig: Maps_ai} displays our results for the two maps.   For the
case without Ceres (panel A) the orbital region of the Astrid family is 
quite stable and regular, with most of the perturbations caused the 
3J:-1S:-1A and 5J:-2A mean-motion resonances.  More interesting is the 
case where Ceres was treated like a massive body (panel B).  As observed by 
\citet{Novakovic_2016}, the linear nodal secular resonance $s-s_C$ now
appears in the region.  Objects whose pericenter frequency is within
$\pm0.3$ arc-sec/yr from $s_C = -59.17$ arc-sec/yr, likely resonators
in the terminology of \citet{Carruba_2009}, are shown as black 
full dots in this figure.  Two other secular resonances involving the
nodal frequency $s_C$ of Ceres are also observed.  Since the difference
for the values of the $g_5$ and $g_7$ precession frequency of the 
pericenter of Jupiter and Uranus is small (4.257 and 3.093 arcsec/yr, 
respectively, which yield a difference of 1.164 arcsec/yr 
\citep{Knezevic_2000}), resonances of resonant argument 
involving $s-s_C$ and combinations of these two frequencies that satisfy 
the D'Alembert rules of permissible arguments are close in proper
element space with respect to the main resonance $s-s_C$.  In this
work we called such resonances ``harmonics'' of the main resonance.
We identified the $s-s_C-g_5+g_7$ and $s-s_C-2(g_5+g_7)$
harmonics, whose likely resonators are shown in Fig.~\ref{fig: Maps_ai} 
as full squares and full hexagons, respectively.

\begin{figure*}
  \centering
  \begin{minipage}[c]{0.50\textwidth}
    \centering \includegraphics[width=3.5in]{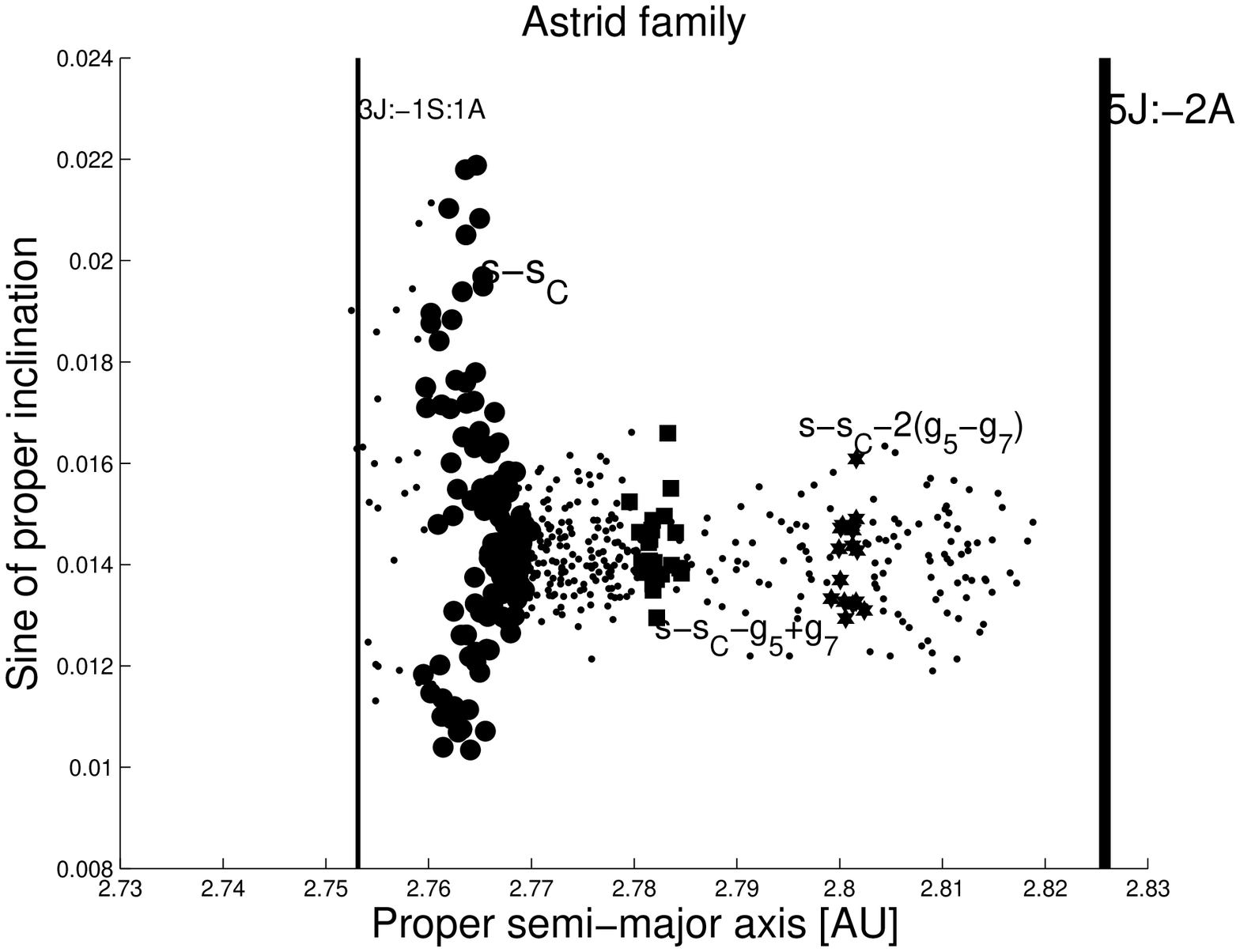}
  \end{minipage}%
  \begin{minipage}[c]{0.50\textwidth}
    \centering \includegraphics[width=3.5in]{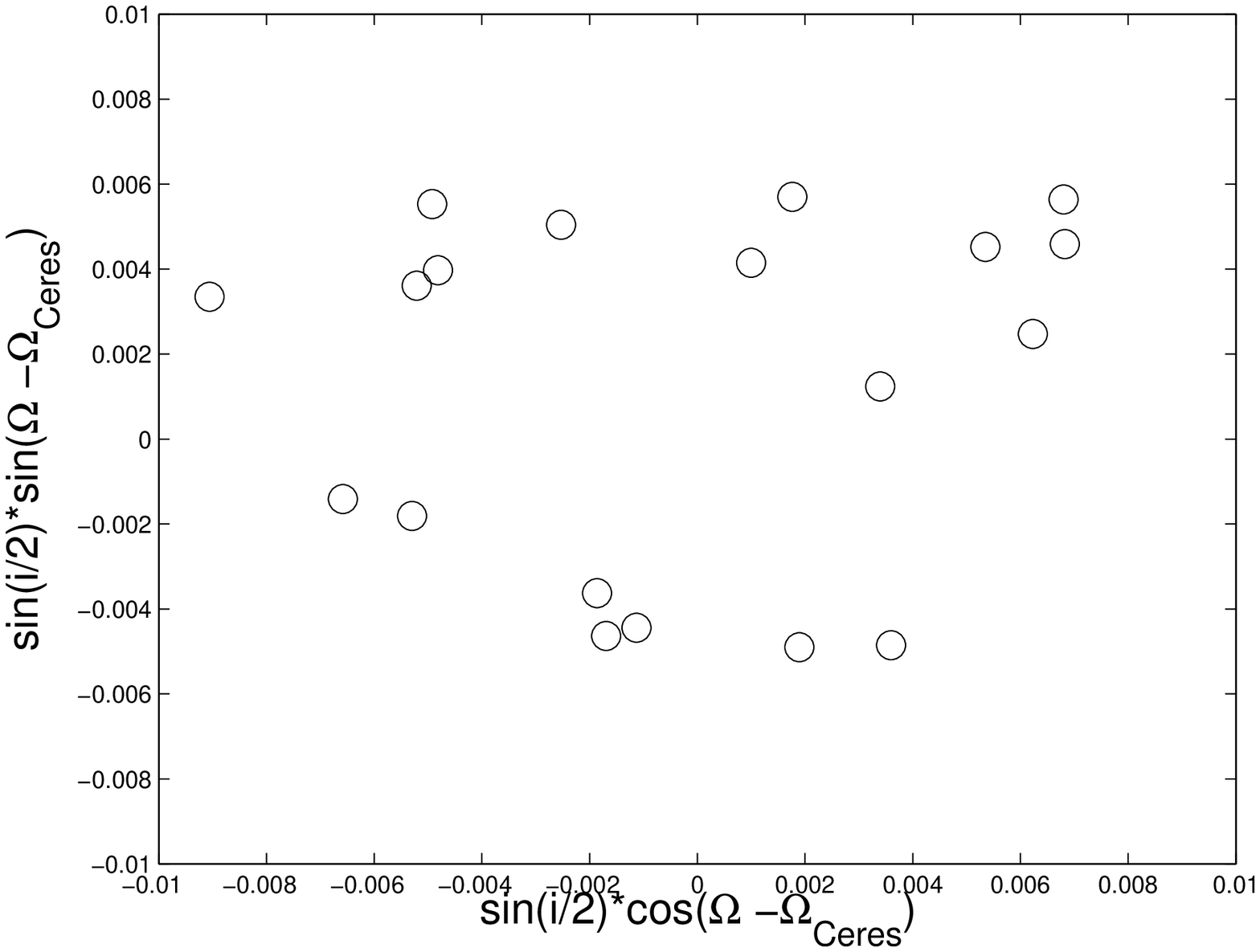}
  \end{minipage}

\caption{An $(a,\sin{(i)})$ projection of 
the 489 HCM Astrid asteroids, with the likely resonators shown in the 
same symbols code as in Fig.~\ref{fig: Maps_ai} (panel A).  
Panel B show a projection in the $(\sin{(i/2)} \cos{(\Omega-{\Omega}_{C})},
\sin{(i/2)} \sin{(\Omega-{\Omega}_{C})}$ of the 19 asteroids 
observed to be in librating states of the $s-s_C$ resonance.} 
\label{fig: Astrid_ssc}
\end{figure*}

To study the resonant dynamics of the Astrid family members, we integrated
the 489 HCM Astrid asteroids with the same scheme used to obtain the
dynamical map in Fig.~\ref{fig: Maps_ai}, panel B.  We then i) 
identify the likely resonators in the $s-s_C$ resonance, and studied
the time evolution of the resonant argument $\Omega -{\Omega}_C$.
We identified 96 likely resonators, and 19 objects (19.8\% of the total) 
whose resonant argument librated around $\pm90^{\circ}$ for 20 Myr, 
the length of the integration.  Unfortunately, the limited number of 
objects in librating states of the $s-s_C$ resonance does not allow to use 
conserved quantities of this resonance to obtain information on the initial 
ejection velocity field, as done by \citet{Vokrouhlicky_2006b} for the Agnia 
family and the $z_1$ secular resonance, or, more recently, by 
\citet{Carruba_2015b} for the Erigone family and the $z_2$ resonance.  
No asteroid was identified in librating states of the $s-s_C-g_5+g_7$, 
$s-s_C-2(g_5+g_7)$, and $s-s_c + g_5-2g_6+g_c$ resonances. 
We then computed proper values of
the resonant frequency $s$, its amplitude $\sin{(i/2)}$, and its
phase $\Omega$ for the 19 resonant objects and Ceres itself.  

Fig.~\ref{fig: Astrid_ssc} displays an $(a,\sin{(i)})$ projection of 
the 489 HCM Astrid asteroids, with the likely resonators shown in the 
same symbol code as in Fig.~\ref{fig: Maps_ai} (panel A).  Panel B show
a projection in the $(\sin{(i/2)} \cos{(\Omega-{\Omega}_{C})},
\sin{(i/2)} \sin{(\Omega-{\Omega}_{C})}$ of the 19 asteroids 
observed to be in librating states of the $s-s_C$ resonance.
One can notice that i), as observed from \citet{Novakovic_2016}, the 
spread in $\sin{(i)}$ of Astrid family members is indeed caused
by the $s-s_C$ nodal resonance, and that, ii) resonant asteroids
seems to oscillate around the stable point at 
$\Omega-{\Omega}_{C} = 0^{\circ}$.  No other stable point was identified
in this work, and the width of the $s-s_C$ resonance is equal 
to $0.8$~arcsec/yr.

To check how fast an initially tight cluster in the 
$(\sin{(i/2)} \cos{(\Omega-{\Omega}_{C})},
\sin{(i/2)} \sin{(\Omega-{\Omega}_{C})}$ would be dispersed beyond recognition,
so losing information about its initial configuration, we followed 
the approach of \citet{Vokrouhlicky_2006b}.  We generated 81 clones
of 183405 2002 YE4, the lowest numbered object in a librating state of the
$s-s_C$ resonance.  The clones are in a 9 by 9 grid in eccentricity 
and inclination, with a step of 0.00001 in eccentricity
and 0.0001 in inclination, and the elements of 183405 as central values
of the grid.   As observed for the $z_2$ resonant asteroids 
in the Erigone family (\citet{Carruba_2015b}, Fig. 9), the initially tight 
cluster becomes uniformly dispersed along the separatrix of the $s-s_C$ 
resonance.  To quantify this effect, we used the polar angle $\Phi$ in the
$(\sin{(i/2)} \cos{(\Omega-{\Omega}_{C})},
\sin{(i/2)} \sin{(\Omega-{\Omega}_{C})}$ plane, as defined in 
\citet{Vokrouhlicky_2006b}.  At each step of the numerical simulation, 
we computed the dispersion $D²_{\Phi}$ in the polar angle $\Phi$ defined
as:

\begin{equation}
D^2_{\Phi}=\frac{1}{N(N-1)}{\displaystyle\sum}_{i \neq j}({\Phi}_i-{\Phi}_j)^2,
\label{eq: D_phi}
\end{equation}

\begin{figure}

  \centering
  \centering \includegraphics [width=0.50\textwidth]{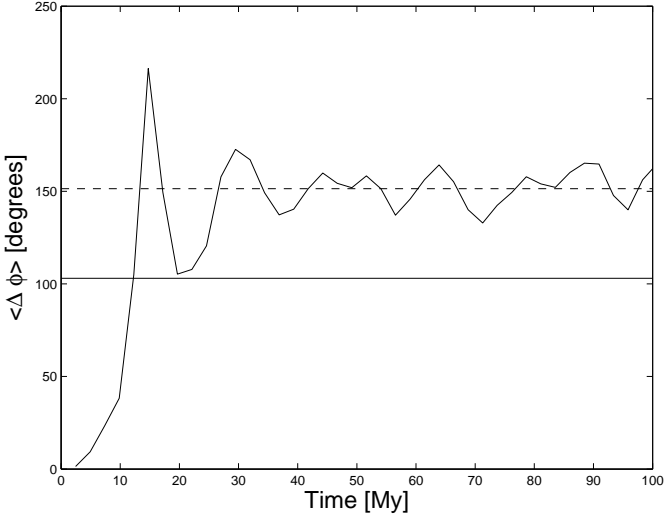}

\caption{Temporal evolution of $D^2_{\Phi}$ of 
Eq.~\ref{eq: D_phi} for the 81 clones of 183405.  The horizontal
black line display the level corresponding to an uniform distribution 
of bodies along a circle \citet{Vokrouhlicky_2006b}.  The dotted line 
display the median value of  $D^2_{\Phi}$ during the simulation.} 
\label{fig: Dphi}
\end{figure}

where N = 81 is the number of integrated bodies and ${\Phi}_i$ is the
polar angle of the $i$-th body (i = 1,...,N).  Since we started with 
a compact cluster, $D^2_{\Phi}$ is initially small ($\simeq 6.61^{\circ}$),
but grows with time because of the differential libration of the bodies
in the resonance (Fig.~\ref{fig: Dphi}).
After only $\simeq$ 12 Myr, i.e., about two libration cycles of the $s-s_c$ 
resonance for 183405, the value of $D^2_{\Phi}$ saturates at $\simeq 103^{\circ}$, 
which corresponds to an uniform distribution of bodies along a circle 
\citet{Vokrouhlicky_2006b}.  This sets a lower limit on the timescale
for dispersion of asteroids in the 
$(\sin{(i/2)} \cos{(\Omega-{\Omega}_{C})},
\sin{(i/2)} \sin{(\Omega-{\Omega}_{C})}$ plane.  Any family that reached
this resonance more than $\simeq$ 12 Myr ago, would have had its members
completely dispersed along the separatrix of the  $s-s_c$ 
resonance, which suggests that Astrid resonant members reached this resonance
more than 12 Myr ago.  

\section{Constraints on terminal ejection velocities from the current
inclination distribution}
\label{sec: inc_constr}

The Astrid family is the product of a relatively recent collision:
\citet{Nesvorny_2015} estimate its age to be $140\pm10$ Myr, 
while \citet{Spoto_2015}, using a V-shape criteria, estimate
the family to be $150\pm32$ Myr old.  Monte Carlo methods 
\citep{Milani_1994, Vokrouhlicky_2006a,
Vokrouhlicky_2006b, Vokrouhlicky_2006c} that simulates the evolution 
of the family caused by the Yarkovsky and YORP effects, where
YORP stands for Yarkovsky-O'Keefe-Radzievskii-Paddack effect, could also be used
to obtain estimates of the age and terminal ejection velocities of the 
family members (these models will be referred as ``Yarko-Yorp'' models 
hereafter).  However, the age estimates from these methods depend on 
key parameters describing the strength of the Yarkovsky force, such as the 
thermal conductivity $K$ and bulk and surface density ${\rho}_{bulk}$ and 
${\rho}_{surf}$, that are in many cases poorly known.  Before attempting our 
own estimate of the family age and terminal ejection velocity field, here 
we analyze what constraints could be obtained
on the possible values of terminal ejection velocities of the 
original Astrid family from its current inclination distribution. 

\begin{figure*}
  \centering
  \begin{minipage}[c]{0.5\textwidth}
    \centering \includegraphics[width=3.5in]{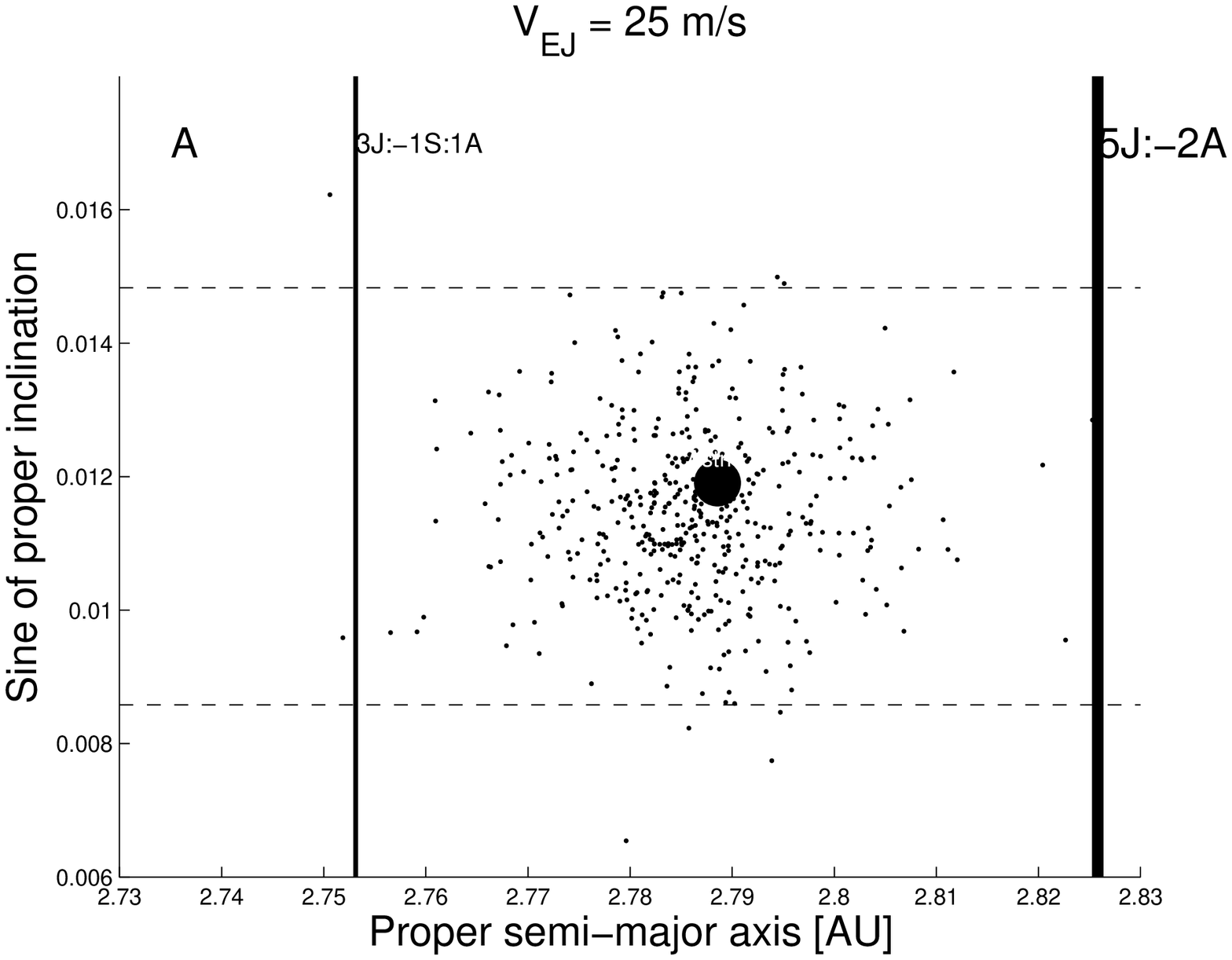}
  \end{minipage}%
  \begin{minipage}[c]{0.5\textwidth}
    \centering \includegraphics[width=3.5in]{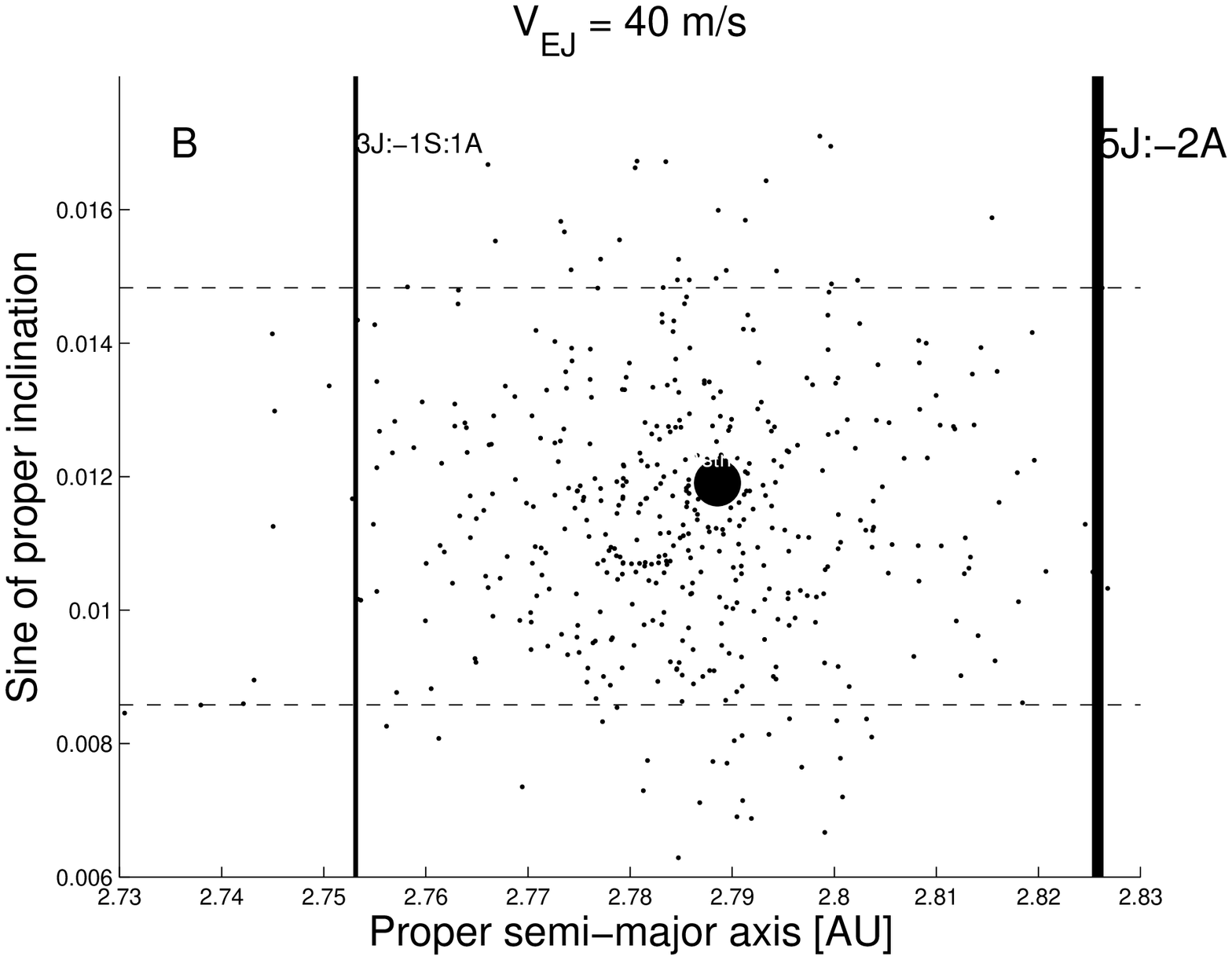}
  \end{minipage}

\caption{An $(a,\sin{(i)}$ projection of the initial orbital dispersion
of a family generated with $V_{EJ} = 25$~m/s (panel A) and $V_{EJ} = 40$~m/s 
(panel B). The full black circle identifies the location of 1128 Astrid 
(that essentially corresponds with the family barycenter), while the 
dashed lines show the minimum and maximum values of $\sin{(i)}$ currently 
observed for members of the Astrid family with $a > 2.77$~au, i.e., those 
that did not yet interacted with the $s-s_C$ secular resonance. The other 
symbols have the same meaning as in Fig.~\ref{fig: Astrid_aei}.}
\label{fig: Sini_constr}
\end{figure*}

In the Yarko-Yorp models, fictitious families are 
generated considering an isotropic velocity field\footnote{Not all 
ejection velocities field are isotropic.  If the fragmentation
was not completely catastrophic, terminal velocities could be rather
anisotropic.  This could actually be the case for the Astrid family,
as also discussed later in this paper.  However, since in this section
we are just interested in setting constraints to the maximum magnitude of the
possible ejection velocity field, we prefer for this purpose 
to use a simpler approach.}, and assuming that 
the fragments are dispersed with a Gaussian distribution  
whose standard deviation follows the relationship:

\begin{equation}
V_{SD}=V_{EJ}\cdot(5km/D),
\label{eq: V_EJ}
\end{equation}

\noindent
where $V_{EJ}$ is the terminal ejection velocity parameter to be estimated,
and $D$ is the asteroid diameter. \citet{Nesvorny_2015} estimated that 
the parent body of the Astrid family was 42.0 km in diameter, which yields
an escape velocity of 33.0 m/s.  Assuming that the $V_{EJ}$ parameter of 
the terminal ejection velocity field would be in the range $0.2 < \beta < 1.5$,
with $\beta = V_{EJ}/V_{esc}$, as observed for most families in the main belt 
\citep{Carruba_2016}, then, expected values of $V_{EJ}$ would be in the 
range from 5 to 50 m/s. If we only consider objects with $a > 2.77$~au, so as 
to eliminate the asteroids that interacted with the $s-s_C$ resonance,
then the currently observed minimum and maximum values of $\sin{(i)}$
of family members are 0.0086 and 0.0148, respectively.  Neglecting possible
changes in $\sin{(i)}$ after the family formation, which is motivated by the
fact that the local dynamics does not seems to particularly affect
asteroids in this region (see Fig.~\ref{fig: Maps_ai}), and will be further
investigated later on, these values set constraints on the possible terminal
ejection velocity parameter $V_{EJ}$ with which the family was created.
Currently, only 7 objects not members of the family are observed at 
sines of inclinations lower that 0.016, i.e., 1.5\% of the current 
number of family members.   We generated synthetic families 
for values of $V_{EJ}$ from 5~m/s up to 40 m/s.  Fig.~\ref{fig: Sini_constr} 
show an $(a,\sin{(i)}$ projection of the initial orbital dispersion of 
the members of the family generated for $V_{EJ} = 25$~m/s (panel A) and 
$V_{EJ} = 40$~m/s.

For $V_{EJ} = 25$ m/s 7 particles (1.5\% of the total) had values of 
$\sin{(i)}$ outside the range of values currently observed, while for 
$V_{EJ} = 40$ m/s these number was 55 (11.5\% of the total).  Based on 
these considerations, it seems unlikely that the ejection velocity parameter
$V_{EJ}$ was larger than $25$~m/s, or a larger number of asteroids outside
the Astrid family at $a > 2.77$~au would be visible today.  This implies
that $\beta = \frac{V_{EJ}}{V_{esc}}$ was at most 0.76, excluding larger 
values associated with more catastrophic events.

\section{Ejection velocities evolution}
\label{sec: term_vel}

\citet{Carruba_2016} recently investigated the shape of the current
distribution of the $v_W$ component of terminal ejection velocity fields
and argued that families that were produced with a $V_{EJ}$ parameter
smaller than the escape velocity from the parent body, are relatively 
young, and are located in dynamically less active regions, as is the 
case of the Astrid family, should be characterized by a leptokurtic
distribution of the $v_W$ component.  This because, assuming that initial
ejection velocities followed a Gaussian distribution, fragments with
initial ejection velocities less than the escape velocity from the 
parent body would not be able to escape.  This would produce a distribution
of ejection velocities more peaked and with larger tails than a Gaussian 
one, i.e., leptokurtic.  While the subsequent dynamical
evolution would tend to cause the distribution of ejection velocities to
be more mesokurtic, this effect would be less intense for families
such Astrid, that are both relatively young and in dynamically less
active regions.

One would therefore expect Astrid to be a relatively leptokurtic family.
However, as also noticed in \citet{Carruba_2016}, the effect of the $s-s_C$
secular resonance tend to increase the dispersion in inclination values
of the family members, and therefore of $v_W$.  While the current
value of ${\gamma}_2$, the parameter associated with the kurtosis
of the $v_W$ distribution (equal to 0 for mesokurtic 
or Gaussian distributions) of the whole Astrid family is quite large, 
( ${\gamma}_2 = 4.43$), if we only consider objects with $a > 2.77$~au that 
did not interacted with the secular resonance, the value of ${\gamma}_2$ is 
just 0.39, more compatible with a relatively somewhat leptokurtic family.
This shows that most of the leptokurtic shape of the currently observed
Astrid family is therefore caused by the interaction of its members with
the $s-s_C$ secular resonance.

To investigate what information the $v_W$ component of the terminal 
ejection velocities could provide on the initial values of the $V_{EJ}$
parameter, we simulated fictitious Astrid families with the currently
observed size-frequency distribution, values of the parameters affecting
the strength of the Yarkovsky force typical of C-type asteroids according to 
\citet{Broz_2013}, i.e., bulk and surface density equal to 
${\rho}_{bulk}={\rho}_{surf} = 1300$~kg/m$^3$, thermal conductivity $K =0.01$~
W/m/K, thermal capacity equal to $C_{th} = 680$~J/kg/K, Bond albedo 
$A_{Bond} =0.02$ and infrared emissivity $\epsilon = 0.9$.
We also generated fictitious families with $V_{EJ} = 5, 10, 15, 20$, and
$25$~m/s, the most likely values of this parameter, according to the 
analysis of the previous section.  Particles were integrated with
$SWIFT\_RMVSY$, the symplectic integrator developed by \citet{Broz_1999}
that simulates the diurnal and seasonal versions of the Yarkovsky effect, 
over 300 Myr and the gravitational influence of all planets plus Ceres.
Values of $v_W$ were then obtained by inverting the third Gauss equation
\citet{Murray_1999}:

\begin{equation}
\delta i = \frac{(1-e^2)^{1/2}}{na} \frac{cos(\omega+f)}{1+e cos(f)} \delta v_W. 
\label{eq: gauss_3}
\end{equation}

where $\delta i= i-i_{ref}$, with $i_{ref}$ the inclination of the barycenter
of the family, and $f$ and $\omega+f$ assumed equal to 30$^{\circ}$ 
and 50.5$^{\circ}$, respectively. Results from \citet{Carruba_2016} show that 
the shape of the $v_W$ distribution is not strongly dependent on the values of 
$f$ and $\omega+f$.

\begin{figure*}
  \centering
  \begin{minipage}[c]{0.50\textwidth}
    \centering \includegraphics[width=3.5in]{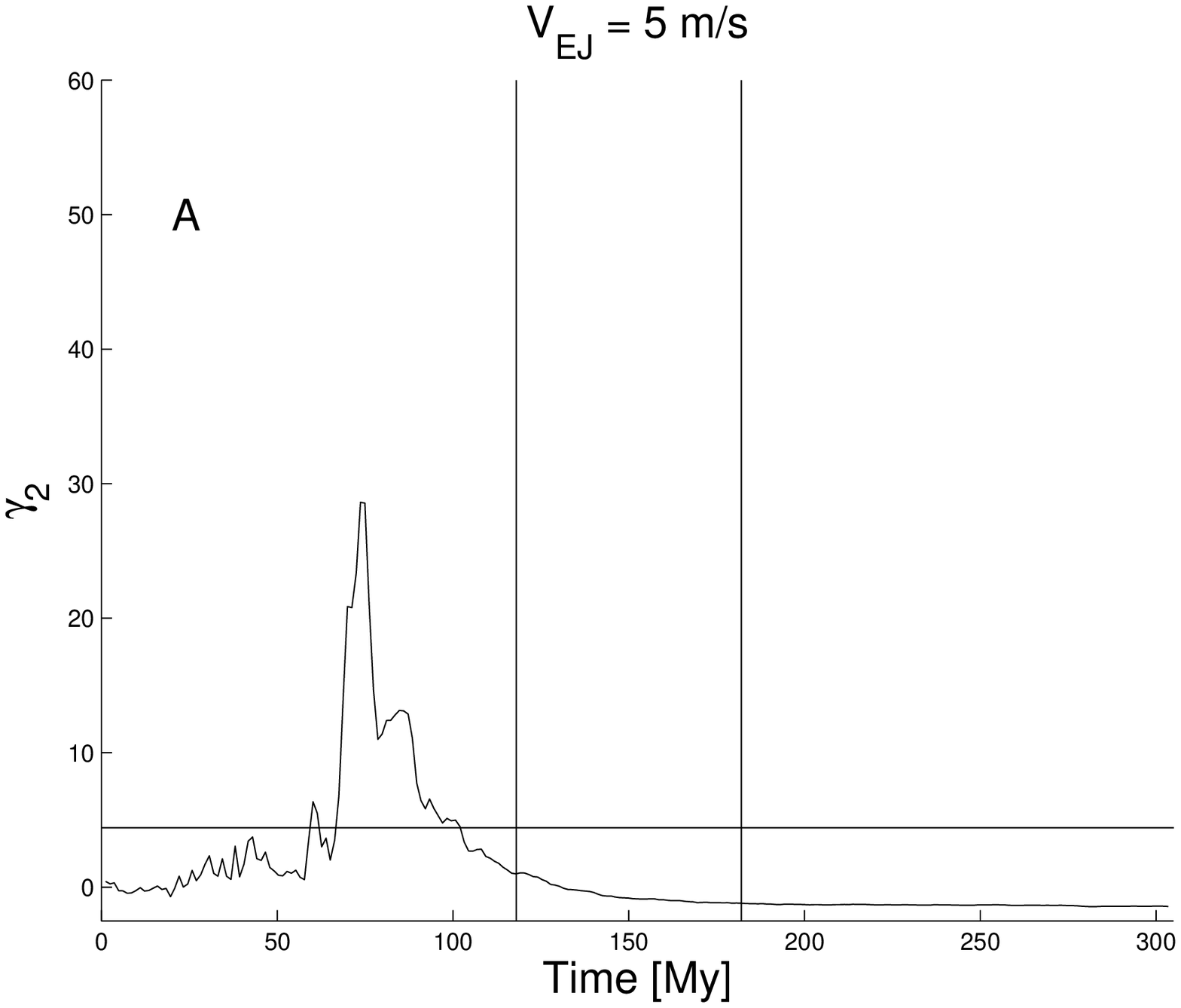}
  \end{minipage}%
  \begin{minipage}[c]{0.50\textwidth}
    \centering \includegraphics[width=3.5in]{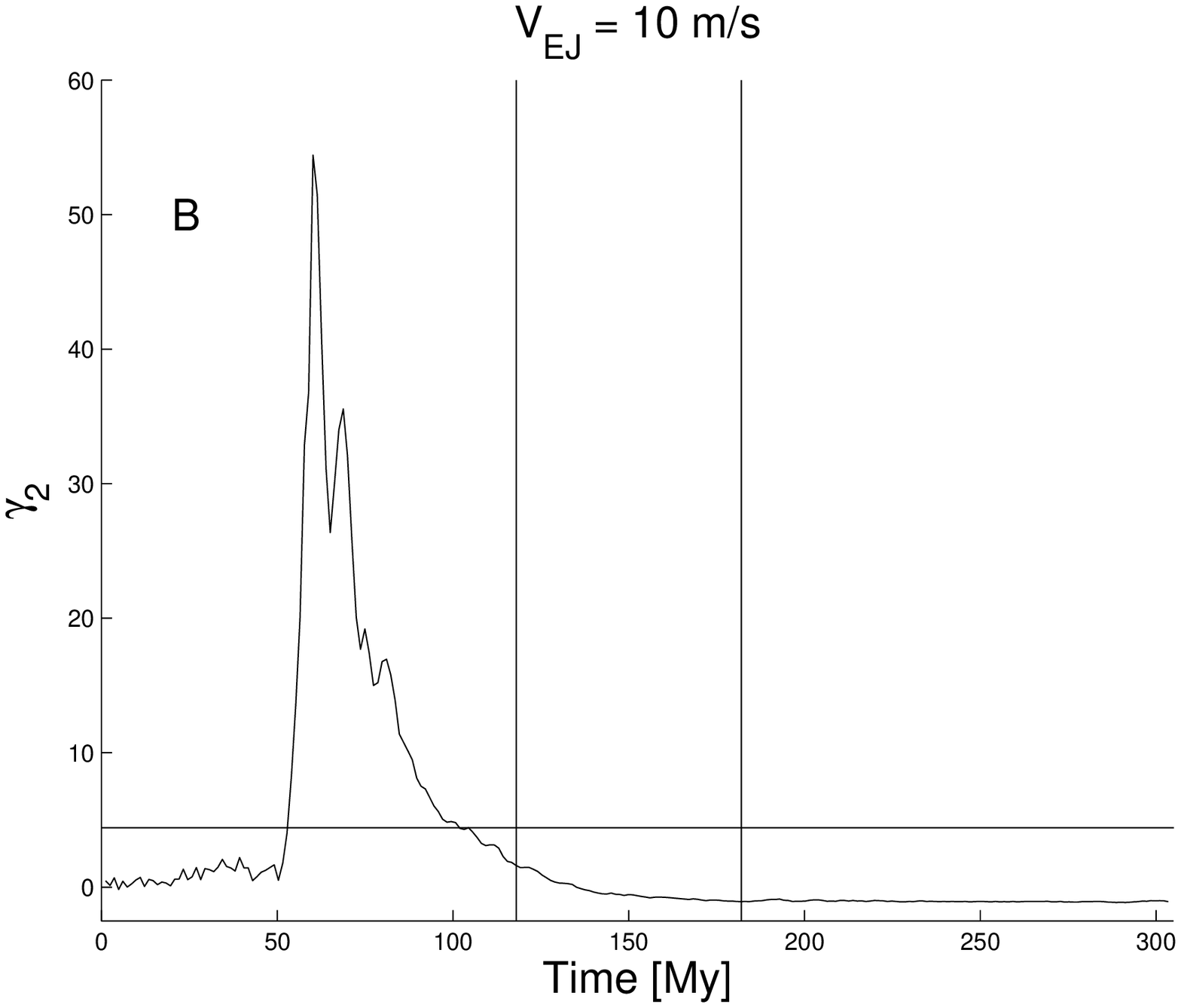}
  \end{minipage}

\caption{Time evolution of the Kurtosis parameter ${\gamma}_2$ for members
of a fictitious family with $V_{EJ} = 5$ m/s (panel A) and 10 m/s (panel B).
The horizontal black line displays the current value of ${\gamma}_2$ for
the real whole Astrid family.  The vertical lines identify the range of
possible ages for the Astrid family, according to \citet{Spoto_2015}.}
\label{fig: gamma2}
\end{figure*}

Fig.~\ref{fig: gamma2} displays the time evolution of the ${\gamma}_2$ 
parameter of the $v_W$ distribution for the fictitious family with 
$V_{EJ} = 5$ m/s (panel A) and 10 m/s (panel B). The peak in the ${\gamma}_2$
value occurs when most particles interacted with the $s-s_C$ secular 
resonance and had their inclination value increased by this resonance.
The current value of ${\gamma}_2$ of the Astrid family is not reached for 
any time inside the range of possible ages, as estimated by \citet{Spoto_2015}
(vertical red lines, the largest range of uncertainty for the age of this
family in the literature. This range of ages corresponds to 
a 1-standard deviation confidence level, obtained by computing a 
Yarkovsky calibration,
with 20\% relative uncertainty, and with an assumed density of 1410~kg/m$^3$),
neither for the simulations with $V_{EJ} = 5$ m/s
nor that with $V_{EJ} = 10$ m/s.  The situation is even worse for families
with larger values of the ejection parameter, for which the peak in 
${\gamma}_2$ is achieved earlier.  This suggests that standard parameters
describing the Yarkovsky force may not apply for the Astrid family.

\begin{figure*}
  \centering
  \begin{minipage}[c]{0.50\textwidth}
    \centering \includegraphics[width=3.5in]{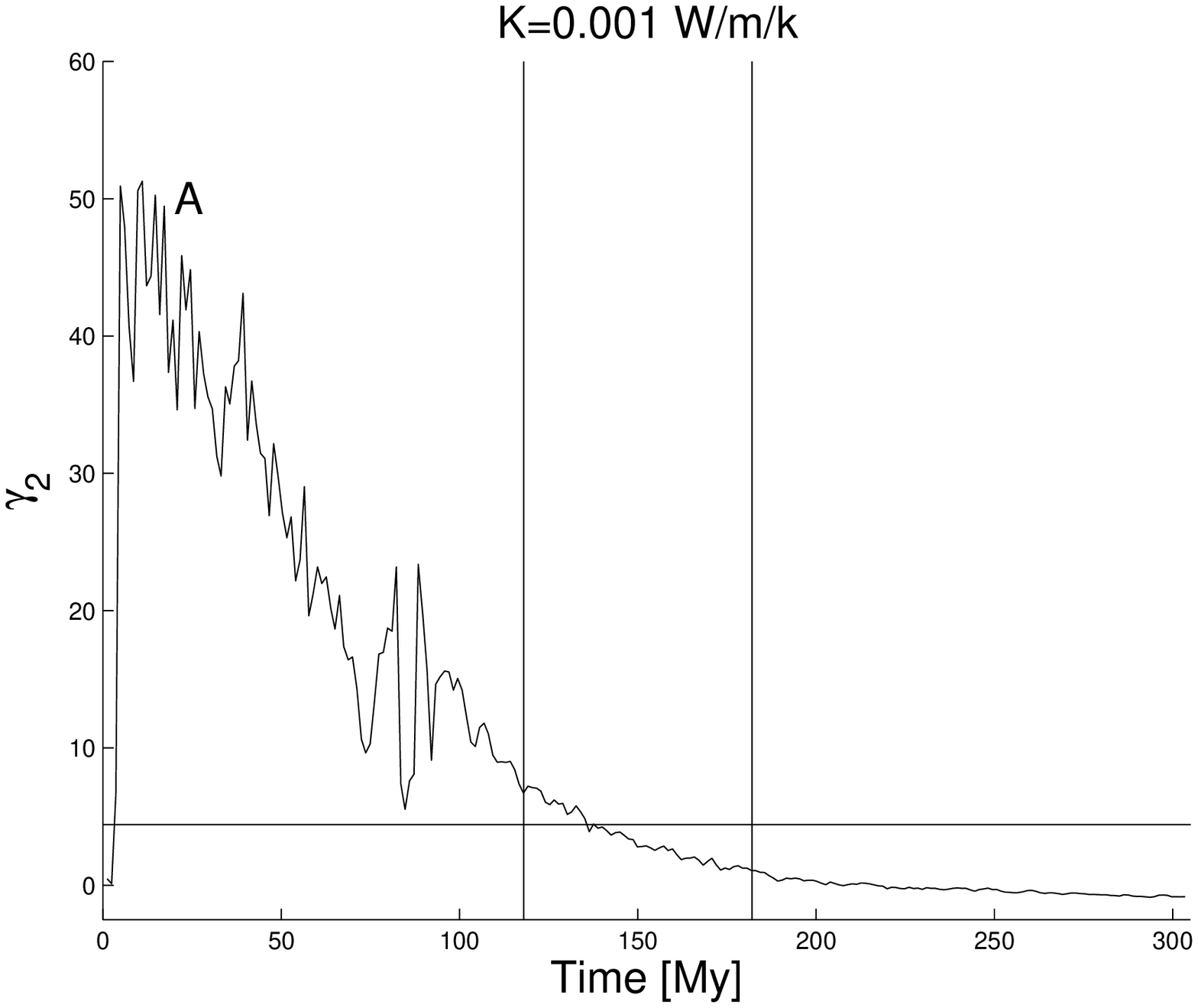}
  \end{minipage}%
  \begin{minipage}[c]{0.50\textwidth}
    \centering \includegraphics[width=3.5in]{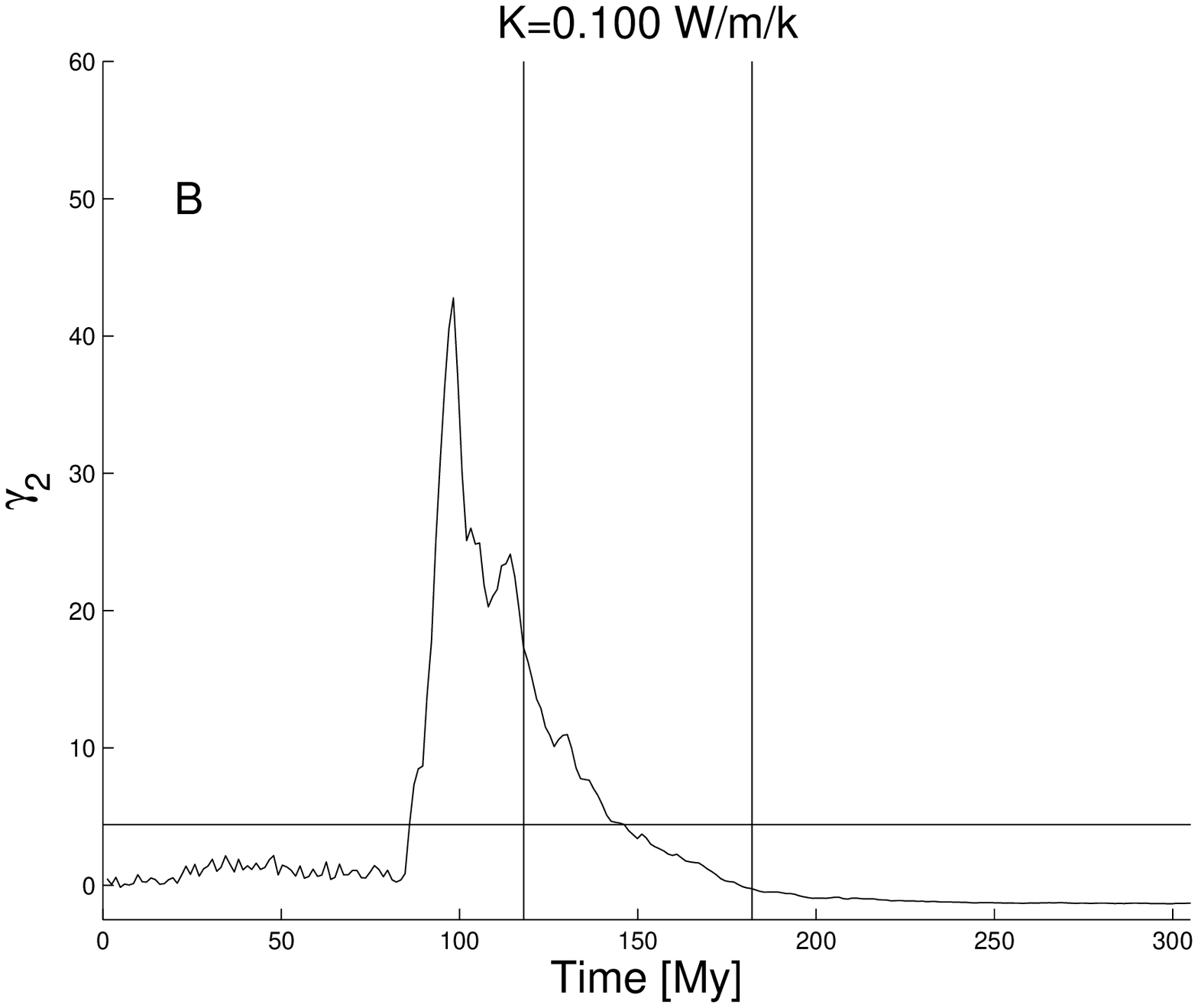}
  \end{minipage}
  \begin{minipage}[c]{0.50\textwidth}
    \centering \includegraphics[width=3.5in]{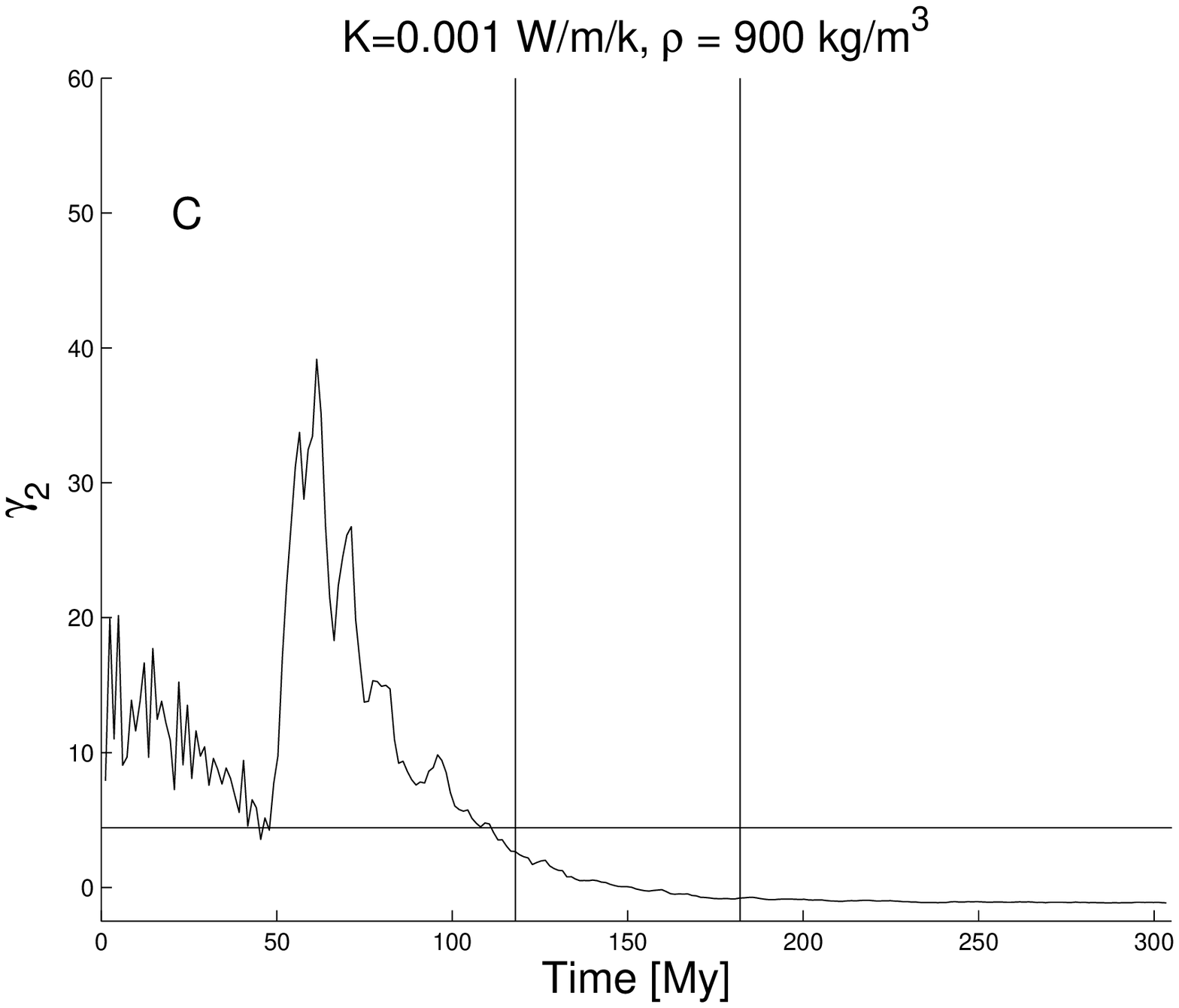}
  \end{minipage}%
  \begin{minipage}[c]{0.50\textwidth}
    \centering \includegraphics[width=3.5in]{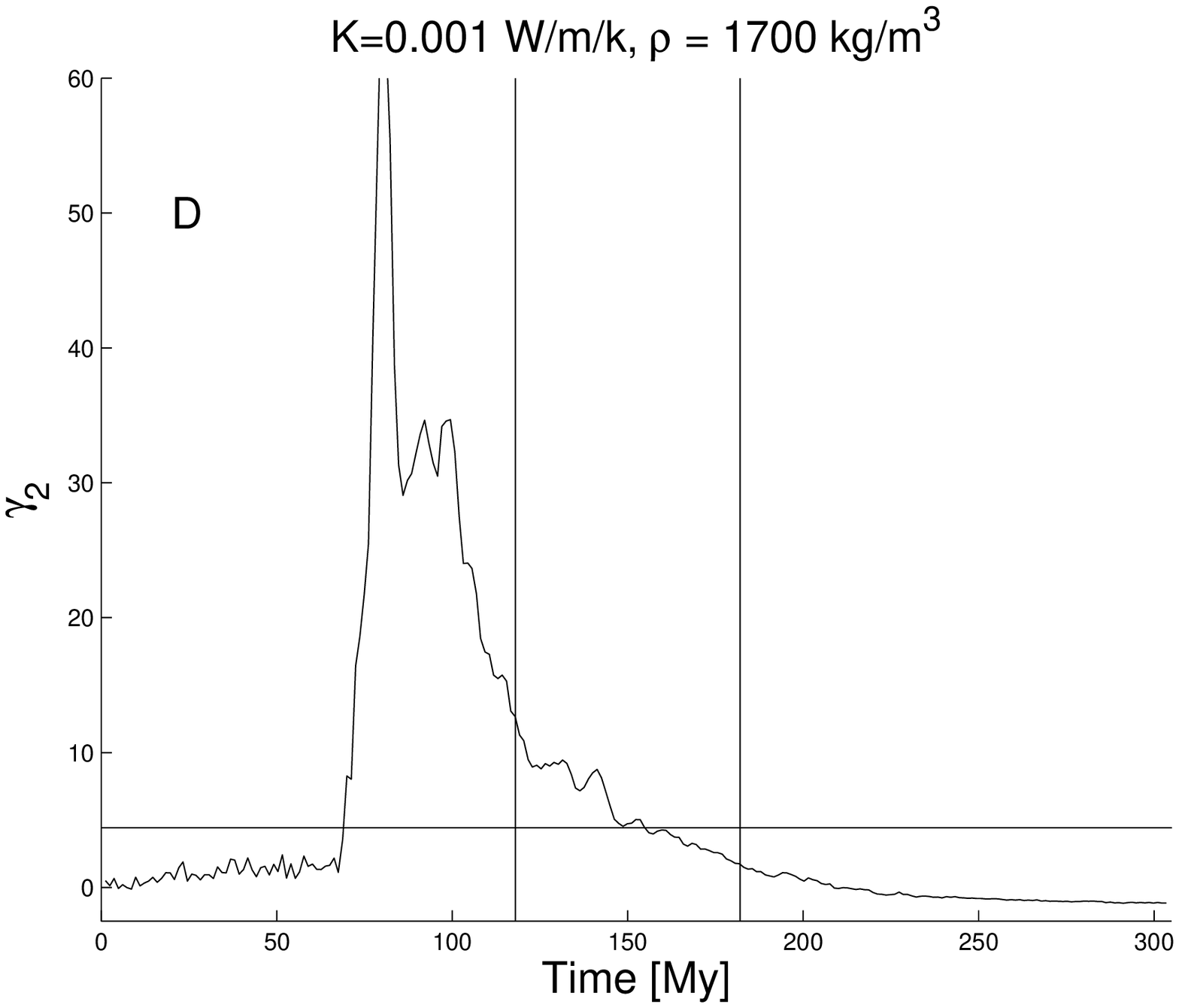}
  \end{minipage}

\caption{Time evolution of the Kurtosis parameter ${\gamma}_2$ for members
of a fictitious family with $V_{EJ} = 10$ m/s and thermal conductivity
$K$ = 0.001 W/m/K (panel A) and 0.100 W/m/k (panel B).  In panel C and D
we display results for $K$ = 0.001 W/m/K and 
${\rho}_{bulk}={\rho}_{surf} = 900$ kg/m$^3$, 
and ${\rho}_{bulk}={\rho}_{surf} = 1700$ kg/m$^3$, respectively.
The symbols have the same meaning as in Fig.~\ref{fig: gamma2}.}
\label{fig: gamma2_K_rho}
\end{figure*}

\citet{Masiero_2012} analyzed the effect that changing the values 
of the Yarkovsky parameters had on estimate of the family age, and found
that the largest effect was associated with changes in the values of
the thermal conductivity and bulk and surface density of asteroids, in 
that order. Based on these results, we first considered two other
possible values of $K$, 0.001 and 0.100 W/m/k, and repeated our simulations
for $V_{EJ} = 10$~m/s.  Results are shown in Fig.~\ref{fig: gamma2_K_rho}.

In both cases, the current value of ${\gamma}_2$ is indeed achieved in
the interval covering the uncertainty associated with Astrid age.    
In the second case, however, the fraction of objects with semi-major axis
lower than 2.7646~au, that crossed the $s-s_c$ resonance, was too small 
at $t=182$ Myr (the maximum possible age for Astrid), when compared
with the current value (15.8\%).  This suggests that $K$ = 0.001 W/m/K
could be closer to the actual value of thermal conductivity of
the real Astrid asteroids.  We then considered the effect of changing 
the bulk and surface density, assumed equal, for simplicity.  We used 
for the two sets of simulations 
${\rho}_{bulk}={\rho}_{surf} = 900$ kg/m$^3$ and 
${\rho}_{bulk}={\rho}_{surf} = 1700$ kg/m$^3$, that are at the extreme of the 
range of values for the density of C-type asteroids \citep{DeMeo_2013}.  
The other parameters were equal
to previous values, and $K$ = 0.001 W/m/K.  Fig.~\ref{fig: gamma2_K_rho}, 
panels C and D, displays our results.  While the values of ${\gamma}_2$
for the first simulation, do not reach the current value in the time 
interval covering the uncertainty associated with Astrid age, larger 
values of the density could be still compatible with our ${\gamma}_2$
test.  Overall, our results suggest that the thermal conductivity
$K$ of Astrid members should be of the order of $K$ = 0.001 W/m/K, while
the mean density of Astrid fragments should be higher than 1000 kg/m$^3$.
Remarkably, results obtained with the ${\gamma}_2(V_W)$ method
are in good agreement with those obtained from independent methods
\citep{Spoto_2015}.

\section{Chronology of the Astrid family}
\label{sec: chron}

\begin{figure*}

  \centering
  \begin{minipage}[c]{0.5\textwidth}
    \centering \includegraphics[width=3.5in]{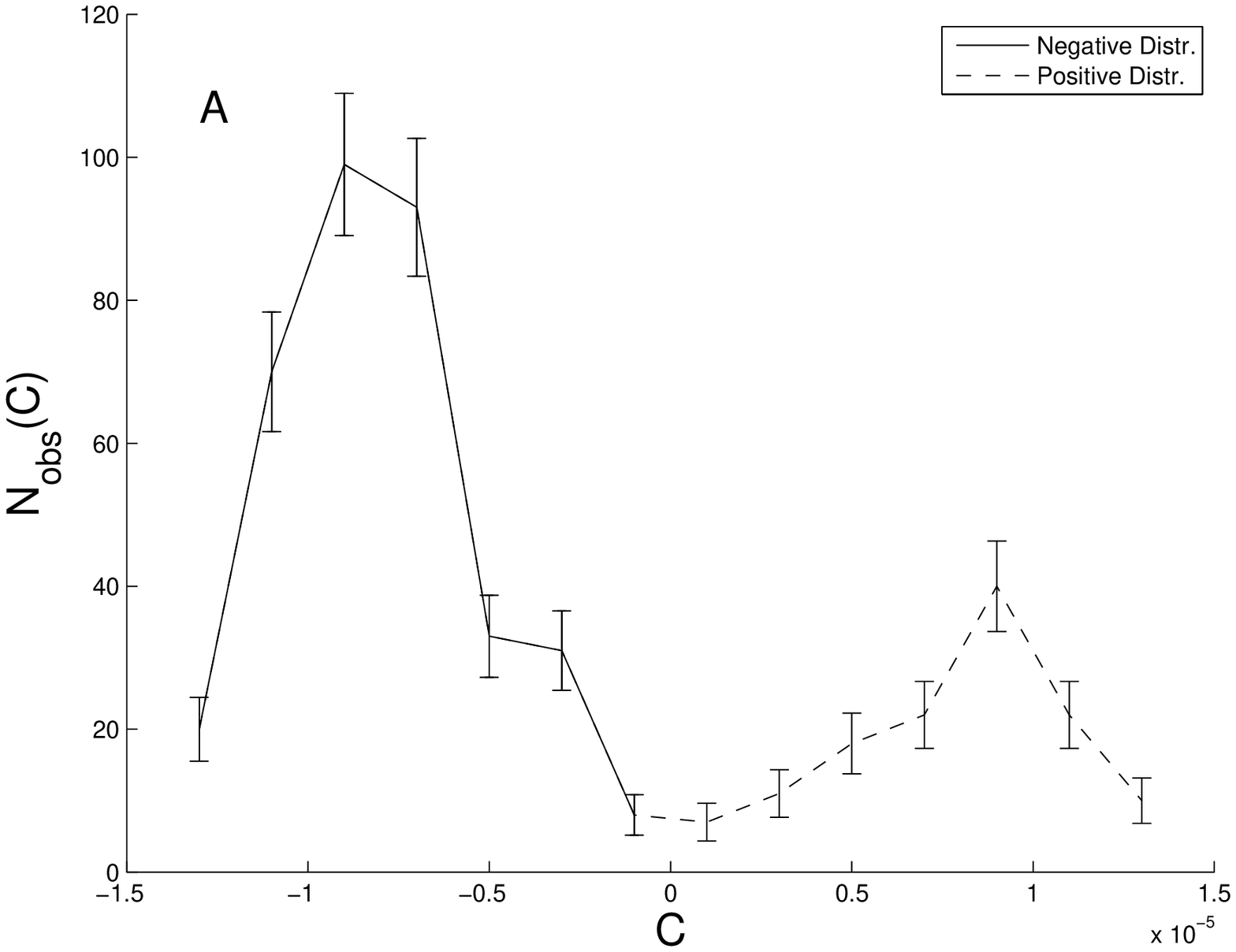}
  \end{minipage}%
  \begin{minipage}[c]{0.5\textwidth}
    \centering \includegraphics[width=3.5in]{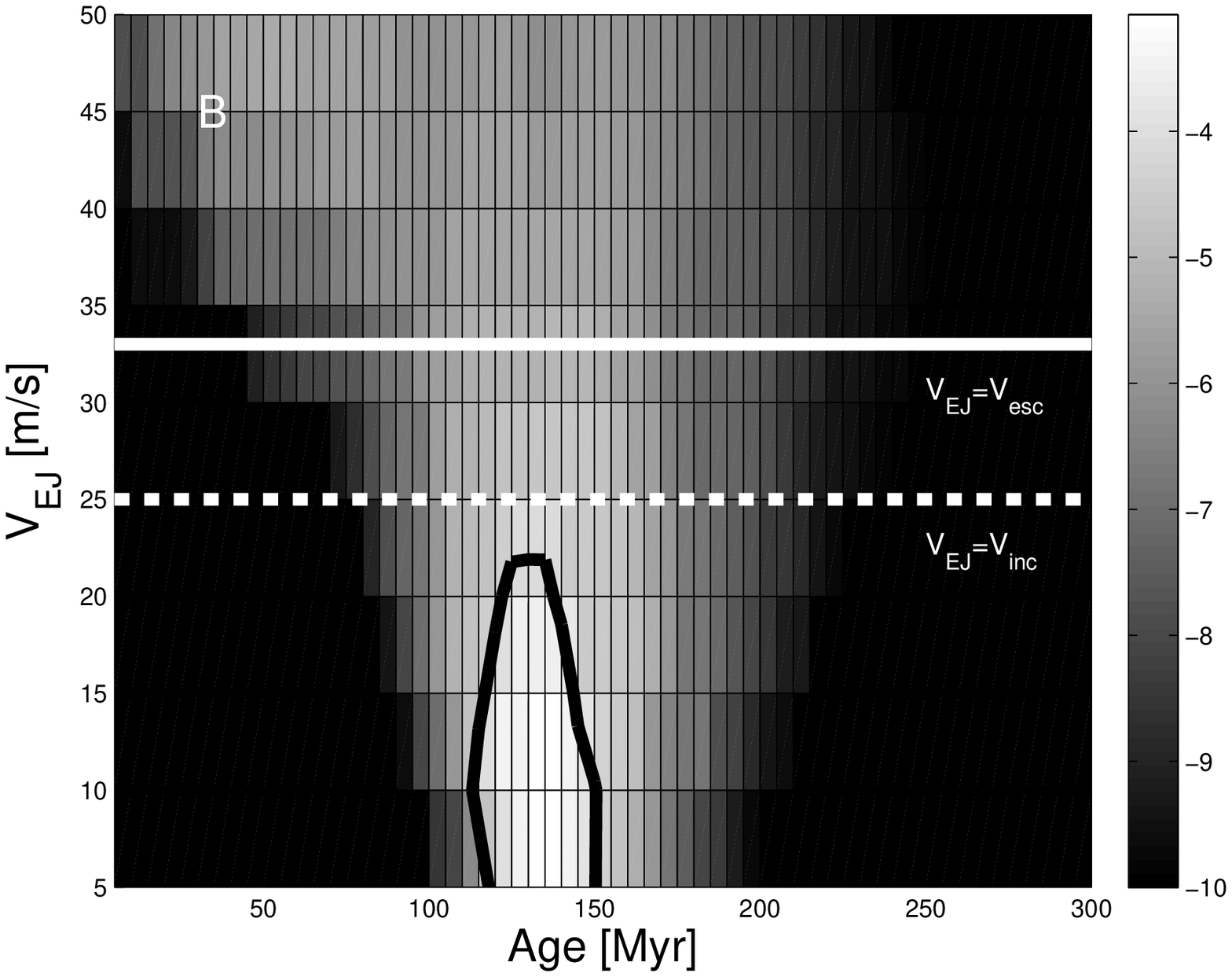}
  \end{minipage}

\caption{Panel A: Histogram of the distribution of $C$ values for the 
Astrid family (blue line).  The dashed line displays the 
positive part of the $C$ distribution.
Panels B: target function ${\psi}_{\Delta C}$ values in ($Age,V_{EJ}$) 
plane for a symmetrical bimodal distribution based on the 
$C$ negative values.  The horizontal full white line display the value of 
the estimated escape velocity from the parent body, 
while the horizontal dashed white line refers to the $V_{EJ} = 25$~m/s limit
obtained from the current inclination distribution in 
Sect.~\ref{sec: inc_constr}.  The black thick
line displays the contour level of ${\psi}_{\Delta C}$ associated
with a 1-sigma probability that the simulated and 
real distribution were compatible.}
\label{fig: Nobs_C_Astrid}
\end{figure*}

Now that the analysis of the current inclination distribution and 
our ${\gamma}_2$ test provided independent constraint on the values 
of the $V_{EJ}$ parameter and of the thermal conductivity and density
of Astrid members, we can try to obtain an independent age estimate
for this family.  We use the approach described in \citet{Carruba_2015a}
that employs a Monte Carlo method \citep{Milani_1994, Vokrouhlicky_2006a,
Vokrouhlicky_2006b, Vokrouhlicky_2006c} to estimate the
age and terminal ejection velocities of the family members.
More details on the method can be found in \citet{Carruba_2015a}.  Essentially,
the semi-major axis distribution of simulated asteroid families is evolved
under the influence of the Yarkovsky effect (both diurnal and seasonal
version), the stochastic YORP force, and changes in values of the past
solar luminosity.  Distributions of a $C$-target function are then
obtained through the equation:

\begin{equation}
0.2H=log_{10}(\Delta a/C),
\label{eq: target_funct_C}
\end{equation}  

where $H$ is the asteroid absolute magnitude, and $\Delta a = a -a_{center}$
is the distance of each asteroid from its family center, here 
defined as the family center of mass.  For the Astrid family this is essentially
equal to the semi-major axis of 1128 Astrid itself.  We can then compare the
simulated $C$-distributions to the observed one by finding the minimum of a 
${\chi}^2$-like function:

\begin{equation}
{\psi}_{\Delta C}=\sum_{\Delta C}\frac{[N(C)-N_{obs}(C)]^2}{N_{obs}(C)},
\label{eq: psi}
\end{equation}

where $N(C)$ is the number of simulated objects in the $i-th$ $C$ interval,
and $N_{obs}(C)$ is the observed number in the same interval.    
Good values of the ${\psi}_{\Delta C}$ function are close to the number of 
the degrees of freedom of the ${\chi}^2$-like variable.  This is given
by the number of intervals in the $C$ minus the number of parameters 
estimated from the distribution (in our case, the the family age and 
$V_{EJ}$ parameter).  Using only intervals with more than 10 asteroids,
to avoid the problems associated with small divisors in Eq.~\ref{eq: psi},
we have in our case 7 intervals for $C < 0$ (see 
Fig.~\ref{fig: Nobs_C_Astrid}, panel A) and 2 estimated parameters, and, 
therefore, 5 degrees of freedom.
If we assume that the ${\psi}_{\Delta C}$ probability 
distribution follows a law given by an incomplete gamma function of arguments 
${\psi}_{\Delta C}$ and the number of degrees of freedom, the value of 
${\psi}_{\Delta C}$ associated with a 1-sigma probability 
(or 68.23\%) of the simulated and real distributions being compatible is 
equal ${\psi}_{\Delta C}=4.3$ \citep{Press_2001}.

The reason why we only considered negative values of $C$ for 
our analysis is that the 
semi-major axis distribution (and, therefore, the $C$ one) is quite
asymmetric. 72.4\% of family members are encountered at lower semi-major
axis than that of 1128 Astrid.  This reflects into a bimodal distribution
of the $C$ values as well, with a more pronounced peak at negative 
$C$ values (see Fig.~\ref{fig: Nobs_C_Astrid}, panel A).   Among the
causes that could have produced this situation, i) the original 
fraction of retrograde rotators produced in the collision 
could have been higher, ii) the ejection velocity field could have
been asymmetrical, with a large fraction of members ejected at lower
semi-major axis, and iii) some of the members of the family at higher
semi-major axis could have been lost in the 5J:-2A mean-motion 
resonance.  Rather than account for any of these mechanisms, or 
better an unknown combination of the three, we preferred in this work to use
a different approach.  Since the most interesting dynamics occurs
for values of semi-major axis lower than the family center, 
we just fitted the distribution of $C$ negative values 
using Eq.~\ref{eq: psi}.  Results of
our simulations are shown in Fig.~\ref{fig: Nobs_C_Astrid}, panel B,
that displays target function ${\psi}_{\Delta C}$ values in the 
($Age,V_{EJ}$) plane.  As determined from the previous section, we used
$K$ = 0.001 W/m/K and ${\rho}_{bulk}={\rho}_{surf} = 1300$ kg/m$^3$.
Values of other parameters of the model such as $C_{YORP}$, ${\delta}_{YORP}$
and $c_{reorient}$ and their description can be found in \citet{Bottke_2015}.

At 1-sigma level, we obtain $T = 135^{+15}_{-20}$ Myr, and 
$V_{EJ}= 5^{+17}_{-5}$~m/s.  Overall, to within the nominal errors, we 
confirmed the age estimates of \citet{Nesvorny_2015} and \citet{Spoto_2015}.  
Independent constraints from Sect.~\ref{sec: inc_constr} imply that 
$V_{EJ} < 25$~m/s, in agreement with our results.

\section{Conclusions}
\label{sec: conc}

Our results could be summarized as follows:

\begin{itemize}

\item We identify the Astrid family in the domain of proper elements,
and eliminated albedo and photometric interlopers.  The Astrid family 
is a C-complex family and C-complex objects dominate the local background.
19 members of the family are in $s-s_C$ resonant librating states, and appear
to oscillate around the stable point at $\Omega-{\Omega}_{C} = \pm 90^{\circ}$.  
The width of the librating region of the $s-s_C$ resonance is equal 
to $0.8$~arcsec/yr, and any cluster of objects injected into the resonance
would have its members completely dispersed along the separatrix of the  
$s-s_c$ resonance on timescales of the order of 10 Myr.

\item Assuming that the original ejection velocity field of the Astrid
family could be approximated as isotropic, the $V_{EJ}$ parameter
describing the standard deviation of terminal ejection velocity
should not have been higher than 25 m/s, or the family would have been more
dispersed in inclination than what currently observed.

\item Interaction with the $s-s_C$ increased the value of the kurtosis
of the distribution of the $v_W$ component of currently observed ejection 
velocities to the large value currently observed (${\gamma}_2 =4.43$).
Simulations of fictitious Astrid families with standard values of 
key parameters describing the strength of the Yarkovsky force for C-type
asteroids, such as the thermal conductivity $K=0.01$ W/m/K, fails to produce
a distribution of asteroids with ${\gamma}_2(v_W)$ equal to the current
value over the possible lifetime of the family.  Constraints from the 
currently observed number of objects that crossed the $s-s_C$ region, 
suggest that $K$ could be closer to 0.001~W/m/K for the Astrid members.
The bulk and surface density should be higher than 1000 kg/m$^3$.

\item Using a Monte Carlo approach to asteroid family determination 
\citep{Bottke_2015, Carruba_2015a}, and values of thermal conductivity 
and asteroid mass density obtained from the ${\gamma}_2(v_W)$ tests, we
estimated the Astrid family to be $T = 135^{+15}_{-20}$~Myr old, and its 
ejection velocity parameter to be in the range $V_{EJ}= 5^{+17}_{-5}$~m/s.
In agreement with what found from constraints from the 
current inclination distribution of family members, 
values of $V_{EJ}$ larger than 25 m/s were not likely to have occurred.
\end{itemize}

Overall, the unique nature of the Astrid family, characterized by its 
interaction with the $s-s_C$ secular resonance and by high values
of the ${\gamma}_2$ parameter describing the kurtosis of the $v_W$ component
of the currently estimated ejection velocity field allowed for the use
of techniques that provided invaluable constraints on the range of 
permissible values of parameters describing the Yarkovsky force, such as the 
surface thermal conductivity and density, not available for other asteroid 
families.

\section*{Acknowledgments}
We are grateful to the reviewer of this paper, Prof. Andrea Milani, for 
comments and suggestions that significantly improved the quality of this paper.
We would like to thank the S\~{a}o Paulo State Science Foundation 
(FAPESP) that supported this work via the grant 14/06762-2, and the
Brazilian National Research Council (CNPq, grant 305453/2011-4).
This publication makes use of data products from the Wide-field 
Infrared Survey Explorer (WISE) and NEOWISE, which are a joint project 
of the University of California, Los Angeles, and the Jet Propulsion 
Laboratory/California Institute of Technology, funded by the National 
Aeronautics and Space Administration.

\bsp

\label{lastpage}


\begin{thebibliography}{}

\bibitem[Beaug\'{e} \& Roig(2001)]{Beauge_2001} Beaug\'{e}, C., 
  Roig, F., 2001, Icarus, 153, 391.

\bibitem[Bendjoya \& Zappal{\`a}(2002)]{Bendjoya_2002} Bendjoya, P., 
Zappal{\`a}, V.\ 2002. Asteroids III, W. F. Bottke Jr., A. Cellino, P. 
Paolicchi, and R. P. Binzel (eds), University of Arizona Press, Tucson, 
613.  

\bibitem[Bottke et al.(2015)]{Bottke_2015} Bottke, W. F., 
and 10 co-authors, 2015, Icarus, 247, 191.

\bibitem[Bro\v{z}(1999)]{Broz_1999} Bro\v{z}, M., 1999.
Thesis, Charles Univ., Prague, Czech Republic.

\bibitem[Bro\v{z} et al.(2013)]{Broz_2013} Bro\v{z}, M., Morbidelli, A., 
 Bottke, W.~F., et~al. 2013, A\&A, 551, A117

\bibitem[Bus \& Binzel(2002a)]{Bus_2002a} Bus, J. S., Binzel, 
R. P. 2002a. Icarus 158, 106.

\bibitem[Bus \& Binzel(2002b)]{Bus_2002b} Bus, J. S., Binzel, 
R. P. 2002b, Icarus 158, 146.

\bibitem[Carruba(2009)]{Carruba_2009} Carruba 2009. MNRAS, 395, 358.

\bibitem[Carruba(2010)]{Carruba_2010} Carruba, V. 2010. MNRAS, 408, 580.

\bibitem[Carruba et al.(2015a)]{Carruba_2015a} Carruba, V., Nesvorn\'{y}, D., 
Aljbaae, S., Domingos, R. C., Huaman, M. E., 2015a, MNRAS, 451, 4763.

\bibitem[Carruba et al.(2015b)]{Carruba_2015b} Carruba, V. Winter, O., Aljbaae, 
S., 2015b, MNRAS, 455, 2279.

\bibitem[Carruba \& Nesvorn\'{y}(2016)]{Carruba_2016} Carruba, V., 
Nesvorn\'{y}, D., 2016, MNRAS, 457, 1332.

\bibitem[DeMeo \& Carry(2013)]{DeMeo_2013} DeMeo, F., 
Carry, B., 2013, Icarus, 226, 723.

\bibitem[Ivezi\'{c} et al.(2001)]{Ivezic_2001} Ivezi\'{c}, \v{Z}, and 
34 co-authors, 2001, AJ, 122, 2749.

\bibitem[Kne\v{z}evi\'{c} and Milani(2000)]{Knezevic_2000} Kne\v{z}evi\'{c}, 
Z., Milani, A. (2000), CMDA, 78, 17.

\bibitem[Lazzaro et al.(2004)]{Lazzaro_2004} Lazzaro, D., 
Angeli, C.A., Carvano, J.M., Moth\'e-Diniz, T., Duffard, R., Florczak, M., 
2004, Icarus 172, 179.

\bibitem[Levison and Duncan(1994)]{Levison_1994} Levison, H. F., Duncan, 
M. J., 1994. Icarus, 108, 18-36.

\bibitem[Masiero et al.(2012)]{Masiero_2012} Masiero, J. R., 
Mainzer, A. K., Grav, T., Bauer, J. M., and Jedicke, R., 2012, APJ, 759, 14. 

\bibitem[Milani \& Farinella(1994)]{Milani_1994} Milani, A.,
Farinella, P., 1994, Nature 370, 40.

\bibitem[Murray and Dermott(1999)]{Murray_1999} Murray, C. D., 
Dermott, S. F., 1999, Solar System Dynamics, Cambridge Univ. Press, Cambridge.

\bibitem[Nesvorn\'{y} et al.(2015)]{Nesvorny_2015} 
Nesvorn\'{y}, D., Bro\v{z}, M., Carruba, V. 2015,   
In Asteroid IV, (P. Michel, F. E. DeMeo, W. Bottke Eds.), Univ. Arizona Press 
and LPI, 297.

\bibitem[Novakovi\'{c} et al.(2015)]{Novakovic_2015}
Novakovi\'{c}, B., Maurel, C., Tsirvoulis, G., Kne\v{z}evi\'{c}, Z.
2015, ApJ, 807, L5.

\bibitem[Novakovic et al.(2016)]{Novakovic_2016} Novakovic, B., 
Tsirvoulis, G., Maro, S., Djosovic, V., \& Maurel, C.\ 2016, arXiv:1601.00905 

\bibitem[Press et al.(2001)]{Press_2001} Press, V.H., 
Teukolsky, S. A., Vetterlink, W. T., Flannery, B. P., 
2001, Numerical Recipes in Fortran 77, Cambridge Univ. Press, Cambridge.

\bibitem[Russell et al.(2015)]{Russell_2015} Russell, C.~T., 
Raymond, C.~A., Nathues, A., et al.\ 2015, IAU General Assembly, 22, 
\#2221738 

\bibitem[Spoto et al.(2015)]{Spoto_2015} Spoto, F., 
Milani, A. Kne\v{z}evi\'{c}, Z. 2015, Icarus, 257, 275.

\bibitem[Tholen et al.(1989)]{Tholen_1989} Tholen, D. J., 
1989, Asteroid Taxonomic Classifications, in Binzel R. P., Gehrels, 
T., Matthews, M. S. (eds), University of Arizona Press, Tucson, 298.

\bibitem[Vokrouhlick\'{y}(1999)]{Vokrouhlicky_1999} Vokrouhlick\'{y}, D.
 1999, A\&A, 334, 362

\bibitem[Vokrouhlick\'{y} et al. (2006a)]{Vokrouhlicky_2006a} 
Vokrouhlick{\'y}, D., Bro{\v z}, M., Morbidelli, A., 
et al.\ 2006a, Icarus, 182, 92. 

\bibitem[Vokrouhlick\'{y} et al. (2006b)]{Vokrouhlicky_2006b} 
Vokrouhlick\'{y} D., Bro\v{z}, M., Bottke, W. F., Nesvorn\'{y}, D., 
Morbidelli, A. 2006b, Icarus, 182, 118.

\bibitem[Vokrouhlick\'{y} et al. (2006c)]{Vokrouhlicky_2006c} 
Vokrouhlick\'{y} D., Bro\v{z}, M., Bottke, W. F., Nesvorn\'{y}, D., 
Morbidelli, A. 2006c, Icarus, 183, 349.
  
\end{thebibliography}
\end{document}